\documentclass[aps,prb,twocolumn,reprint,superscriptaddress,floats,epsfig]{revtex4-2}

\usepackage[colorlinks=true,allcolors=blue]{hyperref}
\usepackage{graphicx}
\usepackage{dcolumn}
\usepackage{orcidlink}
\usepackage{csquotes}
\usepackage{float}
\usepackage{amsmath}
\usepackage{amssymb}
\usepackage{amsbsy}
\usepackage{ulem}
\usepackage{bm}
\usepackage{braket}
\usepackage{xcolor}
\usepackage{comment}
\usepackage{subfigure}

\newcommand{\bib}{\bibitem}
\newcommand{\beq}{\begin{equation}}
\newcommand{\eeq}{\end{equation}}
\newcommand{\bea}{\begin{eqnarray}}
\newcommand{\eea}{\end{eqnarray}}
\newcommand{\al}{\alpha}
\newcommand{\de}{\delta}

\newcommand{\ep}{\epsilon}
\newcommand{\ga}{\gamma}

\newcommand{\lam}{\lambda}
\newcommand{\si}{\sigma}
\newcommand{\ta}{\theta}
\newcommand{\om}{\omega}
\newcommand{\la}{\langle}
\newcommand{\ra}{\rangle}
\newcommand{\dg}{\dagger}
\newcommand{\non}{\nonumber}

\newcommand{\red}{\textcolor{red}}
\usepackage{tikz}
\usetikzlibrary{positioning}

\begin{document}




\author{Maitri Ganguli\,\orcidlink{0009-0009-4701-2459}}
\affiliation{Department of Physics, Indian Institute of Science, Bengaluru 560094, India}


\author{Diptiman Sen\,\orcidlink{0000-0002-6926-9230}}
\affiliation{Center for High Energy Physics, Indian Institute of Science, Bengaluru 560094, India}

\title{Su-Schrieffer-Heeger model driven by 
sequences of two unitaries: periodic, 
quasiperiodic, aperiodic and random protocols}




\begin{abstract}
We study the effect of driving the Su-Schrieffer-Heeger model using two unitary operators $U_1$ and
$U_2$ in different combinations; the unitaries differ in the
values of the inter-cell hopping amplitudes. Specifically, 
we study the cases where the unitaries are applied periodically, 
quasiperiodically, aperiodically and randomly. For a periodic
protocol, when $U_1 = e^{-i H_1 T/2}$ and $U_2 = e^{-i H_2 T/2}$
are applied alternately, we find that end modes may appear, but the number of 
end modes does not always agree with the winding number which is
a $Z$-valued topological invariant. We then study
the Loschmidt amplitude ($LA$) starting with a initial state
which is an end mode of $H_1$.
We find that the $LA$ exhibits pronounced oscillations whose
Fourier transform has a peak at a frequency which is equal to the quasienergy of an end mode of $U$. Next, 
when $U_1$ and $U_2$ are applied in a quasiperiodic or
aperiodic way (we consider the
Fibonacci and Thue-Morse protocols as examples), we
study the Loschmidt echo ($LE$) starting with an initial state which is an 
end mode of the Hamiltonian $H_1$. When the inter-cell
hoppings differ by a small amount denoted by $\epsilon$, and the
time period $T$ of each unitary is also small, the
distance between the unitaries is found to be proportional
to $\epsilon T$. We then find
that the $LE$ oscillates around a particular value for a 
very long time before decaying to zero. The deviation of the
value of the $LE$ from 1 scales
as $\epsilon^2$ for a fixed value of $T$, while the time after
which the $LE$ starts decaying to zero has an interesting 
dependence on $\epsilon$ and $T$. Finally, when $U_1$ and
$U_2$ are applied in a random order, the $LE$ rapidly decays 
to zero with increasing time. We have presented a qualitative
understanding of the above results.
\end{abstract}


\maketitle

\section{Introduction}
\label{sec1}

Topological phases
of matter are characterized by gapped bulk states and
boundary states whose energies lie within the bulk
gap~\cite{fu07,hasan10,qi11,shen12,bernevig13,asboth16}. 
A characteristic feature of such systems is the existence
of a bulk-boundary correspondence. Namely, there are 
topological invariants derived from the bulk bands which
count the number of boundary modes. For instance,
one-dimensional systems may have a $Z$-valued
topological invariant called the winding number 
which counts the number of modes at each end of an open
system~\cite{niu12,wade13}.

In parallel, periodically driven systems have been studied 
extensively for several years because of the wide variety 
of unusual phenomena that they can exhibit~\cite{blanes09,shev10,ales13,bukov15,ales16,
mika16,oka19,band21,sen21}. In particular, periodic 
driving can be used to engineer topological phases of matter~\cite{kita10,kita11,lindner11,kundu14,dora12,tong13,thakur13,thakur14,katan13,rudner13,nathan15,carp15,thakur17,mukh16,mukh18,zhou18,bandyo22,agrawal22,ghuneim25}, generate
Floquet time crystals~\cite{khemani16,else16,zhang17} and other novel steady states~\cite{russo12,laza14}, produce dynamical localization~\cite{nag14}, dynamical
freezing~\cite{das10,mondal12,iubini19}, and other dynamical transitions~\cite{heyl13,sen16,heyl18,karr13,kriel14,andra14,
canovi14,sharma16,sarkar20,arze20,aditya22},
tune a system into ergodic or nonergodic phases~\cite{mukh20,mukh22},
and generate emergent conservation laws~\cite{haldar21}.
The end modes of a one-dimensional topological system
generated by periodic driving are observable through transport measurements~\cite{kundu13,sur21}.

Apart from periodic driving, there have been several 
studies of the effects of quasiperiodic, aperiodic and 
random driving~\cite{roosz16,nandy17,nandy18,maity19,mukh20b,zhao22}.
These studies have generally studied the effects of the
driving on the properties of the bulk, such as the
thermalization of the system under time evolution starting
from a given initial state. However, the effects of 
quasiperiodic, aperiodic and random driving on the end modes of a 
topological system have not been studied in as much 
detail until now (see, however, a recent study of the 
transverse field Ising model under Fibonacci driving~\cite{schmid25}).
This will be the focus of the work reported here.

The plan of this paper is as follows. In Sec.~\ref{sec2}, we will briefly review the topological properties of the 
Su-Schrieffer-Heeger (SSH) model. Then in Sec.~\ref{sec3}, we
will discuss two
different unitary operators $U_1 = e^{-i H_1 T/2}$ and $U_2
= e^{-i H_2 T/2}$ that we will use 
to drive the system in different ways.
We consider the effect of
periodic driving in which the unitaries are combined as $U_2 U_1$
which has a time period $T$, and this is
then performed repeatedly. After discussing the eigenspectra of 
$U_1$ and $U_2$ separately, we discuss the eigenspectrum of $U_2
U_1$ and whether this product operator has any end modes.
We study if the number of end modes agree with the winding
number which is a topological invariant.
Next, we look at the Loschmidt amplitude ($LA$) starting with 
an end mode of $H_1$, denoted 
as $| \phi (0) \ra$. Namely, we calculate the time-evolved state 
$| \phi (t) \ra = (U_2 U_1)^n | \phi (0) \ra$, where $t=nT$, and 
then compute $LA (n) = \langle \phi (t) | \phi (0) \rangle$. 
We find
that the $LA$ exhibits pronounced oscillations. We therefore
look at the Fourier transform of the $LA$ and find some peaks;
we are able to explain the most prominent peak in terms of 
one of the gaps in the quasienergy spectrum of $U_2 U_1$.
In Sec.~\ref{sec4}, we consider a driving protocol in
which the $U_1$'s and $U_2$'s form a quasiperiodic Fibonacci sequence.
We study the Loschmidt echo ($LE$), defined as $|\langle \phi (t) | \phi (0) \rangle|^2$, starting with the eigenmode of $U_1$ which
is localized near the left end of the system.
If $U_1$ and $U_2$ are close to each other, We find that the $LE$
oscillates about a mean value which remains close to 1 for a surprisingly long time. The deviation of the mean value from 1
scales quadratically with the distance between $U_2$ and $U_1$.
The oscillation frequency is found to be equal to
one of the prominent gaps in the quasienergy spectrum of $U_1$.
Eventually, after a very long time denoted as $T_p$, the $LE$ starts decaying towards zero, and we study the dependence of 
$T_p$ on the parameters of the model. In Sec.~\ref{sec5},
we study a driving protocol in which $U_1$ and $U_2$
form an aperiodic Thue-Morse sequence.
In Sec.~\ref{sec6}, we study what happens when $U_1$ and $U_2$ 
act in a random order on the left end mode of $H_1$.
We discover that the $LE$ decays towards zero instead of
oscillating about some mean value.
(In Secs. \ref{sec4}-\ref{sec6}, we take $U_1 = 
e^{- i H_1 T}$ and $U_2 = e^{- i H_2 T}$ respectively).
In Sec.~\ref{sec7}, we summarize our
main results and point out some directions for future studies.

\section{Su-Schrieffer-Heeger Model}
\label{sec2}

\begin{figure}[H]
\centering
\includegraphics[width=0.8\hsize]{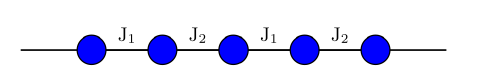}
\caption{Schematic picture of the hopping amplitudes of the SSH model. There are two types of hopping
denoted as $J_1$ (intra-cell hopping) and $J_2$ 
(inter-cell hopping).}
\label{fig1}
\end{figure}

In this section we will briefly review a well-known 
one-dimensional topological system known as the SSH model.
The SSH model consists of a one-dimensional chain where 
each unit cell consists of two sites, 
and the nearest-neighbor hopping amplitudes are different
within a unit cell and between two unit cells. For
a system with $L$ sites (we will assume that $L$ is even)
and open boundary conditions, the Hamiltonian is given by
\bea H &=& J_1 ~\sum_{j=1}^{L/2} ~(a_j^\dg b_j +
b_j^\dg a_j) \non \\
&& +~ J_2 ~\sum_{j=1}^{L/2-1} ~(b_j^\dg 
a_{j+1} + a_{j+1}^\dg b_j), \label{ham1} \eea
where $j$ is the unit cell label, $a_j, b_j$ denote the particle 
annihilation and creation operators respectively at the left
and right sites of the unit cell labeled $j$. (We will take 
the particles to be spinless fermions). We will assume for
simplicity that the intra-cell and inter-cell hopping 
amplitudes, $J_1$ and $J_2$, are both positive.
A schematic picture of the model is shown in Fig.~\ref{fig1}.

It is known that the model described in Eq.~\eqref{ham1} has two
energy bands in the bulk, and the bands are separated by a gap if $J_1 \ne J_2$. The model is in a topological 
phase if $J_1 < J_2$ and in a non-topological phase if 
$J_1 > J_2$; the two phases are 
separated by a quantum phase transition at $J_1 = J_2$ where 
the bulk is gapless. If $J_1 < J_2$, there is a
zero-energy state localized near each end of the system.
In the limit $L \to \infty$, the left-localized end
mode has a normalized wave function of the form
\bea \psi_L (a,j) &=& \sqrt{1 - \frac{J_1^2}{J_2^2}} ~\left( - ~\frac{J_1}{J_2} \right)^{j-1}, \non \\ 
\psi_L (b,j) &=& 0, \label{endmode} \eea
where $j=1,2,3,\cdots$.
If $J_1 \ge J_2$, there are no end modes.

It will be useful later to find the overlap between the 
end modes of two SSH models with different values of the
ratio $J_1/J_2$. Let the values of $J_1/J_2$ in the
two models be given by $\lam_1$ and $\lam_2$ respectively,
both assumed to be smaller than 1.
Using Eq.~\eqref{endmode}, we find that the desired overlap between the wave functions of the two end modes, 
$\psi_{L1}$ and $\psi_{L2}$, is given by
\beq \la \psi_{L1} | \psi_{L2} \ra ~=~ \frac{\sqrt{(1 - \lam_1^2) 
(1 - \lam_2^2)}}{1 - \lam_1 \lam_2}. \label{over1} \eeq
If $\lam_1, \lam_2$ are close to each other, so that
$\lam_1/\lam_2 = 1 - \ep$ where $\ep \ll 1$,
we find that 
\beq \la \psi_{L1} | \psi_{L2} \ra ~=~ 1 ~-~ \frac{\ep^2}{2} ~\frac{\lam_2^2}{(1- \lam_2^2)^2} \label{over2} \eeq
plus terms of higher order in $\ep$. We thus see that the
overlap differs from 1 by a term of order $\ep^2$ 
if $1 - \lam_1/\lam_2$ is of order $\ep$.

The topological and non-topological phases of the SSH
model are distinguished by a
topological invariant called a winding number. This is defined as
follows. We write the Hamiltonian in Eq.~\eqref{ham1} in momentum
space as
\bea H &=& \sum_{k=-\pi}^\pi ~\left( \begin{array}{cc}
a_k^\dg & b_k^\dg \end{array} \right) ~H_k ~\left( \begin{array}{c}
a_k \\ 
b_k \end{array} \right), \non \\
H_k &=& \left( \begin{array}{cc}
0 & J_1 + J_2 e^{-ik} \\
J_1 + J_2 e^{ik} & 0 \end{array} \right). \label{ham2}
\eea
The energy bands are therefore given by
\beq E_{k \pm} ~=~ \pm \sqrt{J_1^2 + J_2^2 + 2 J_1 J_2 \cos k},
\label{disp} \eeq
which are separated from each other by a gap given by
$2 | J_1 - J_2|$. The $k$-dependent Hamiltonian can be 
written in terms of two Pauli matrices $\si^x, ~\si^y$ as
\bea H_k &=& a_{k,x} ~\si^x ~+~ a_{k,y} ~\si^y, \non \\
a_{k,x} &=& J_1 + J_2 \cos k, ~~~~a_{k,y} ~=~ J_2 \sin k. 
\label{hk} \eea
We now think of $(a_{k,x},a_{k,y})$ as the coordinates of a point in
a two-dimensional plane. The number of times this point winds around
the origin $(0,0)$ is the winding number $W$. (In general,
this can be any integer, hence $W$ is called a $Z$-valued
topological invariant). To calculate this 
number, we consider the angle made by the point with respect 
to the $x$-axis as
\beq \phi_k ~=~ \tan^{-1} \left( \frac{a_{k,y}}{a_{k,x}} \right).
\eeq
Then the winding number is given by
\bea W &=& \frac{1}{2\pi}~ \int_{-\pi}^{\pi} dk ~\frac{d \phi_k}{dk}
\non \\
&=& \frac{1}{2\pi}~ \int_{-\pi}^{\pi} dk ~\frac{a_{k,x} 
\frac{d a_{k,y}}{dk} ~-~ a_{k,y} \frac{d a_{k,x}}{dk}}{a_{k,x}^2 ~+~ 
a_{k,y}^2}. \label{wind} \eea
We find that $W=1$ in the topological phase with $J_1 < J_2$, 
while $W=0$ in the non-topological phase with $J_1 > J_2$.
$W$ is not defined at the transition $J_1 = J_2$ since the 
point passes through the origin in that case, i.e., $(a_{k,x},a_{k,y}) 
= (0,0)$ for $k = \pi$. 

We note that if we modify the model so that the Hamiltonian
has a term like $\de J \sum_j (a_j^\dg a_j - b_j^\dg b_j)$, then
$H_k$ would have a term proportional to the third Pauli matrix,
namely, $\de J~ \si^z$. Then the point with coordinates 
$(a_{k,x},a_{k,y},
a_{k,z})$ will move in three dimensions as $k$ varies, and 
it would not be possible to define a winding number.

\section{Periodic driving with two unitaries}
\label{sec3}

We will now begin our studies of the SSH model driven using different
protocols. Our main aim will be see what effects these have on the
end modes. In this section, we will examine what happens when two 
different time-evolution unitaries are applied alternately,
in particular, whether this generates some end modes which are not
present when only one of the operators is applied. To this end, we
consider two Hamiltonians, $H_1$ and $H_2$, which are both
of the SSH form. For $H_1$, we take the hoppings to be $J_1$ 
and $J_2 = J' + \al$, and for $H_2$, the hoppings are
taken to be $J_1$ and $J_2 = J' - \al$; the values
of $J_1$, $J'$ and $\al$ will be specified below. Thus the
intra-cell hopping $J_1$ will be held fixed while the 
inter-cell hoppings will alternate between two values $J' \pm \al$.
The Hamiltonians $H_1$ and $H_2$ will each be
applied for a time equal to $T/2$, so that the
time-evolution operators for the two half cycles are given by
the unitaries $U_1 = e^{-i H_1 T/2}$ and $U_2 = e^{-i H_2 T/2}$. 
(We will set $\hbar = 1$ throughout this paper).
The time-evolution operator $U$ for one full cycle with
time period $T$ is then given by
\beq U(T) ~=~ U_2 U_1 ~=~ e^{-i H_2 T/2} ~e^{-i H_1 T/2}.
\label{u2} \eeq

Given a Floquet operator $U(T)$, it is convenient to define a
time-independent Floquet Hamiltonian $H_F$ through the relation
\beq U(T) ~=~ e^{-i H_F T}. \label{hf} \eeq
Since our system consists of non-interacting particles, we can
write $U = \prod_k U_k$ and $H_F = \sum_k H_{F,k}$. We have
\beq U_k ~=~ e^{-i H_{2,k} T/2} ~e^{-i H_{1,k} T/2} ~=~
e^{-i H_{F,k} T}, \label{uk1} \eeq
where $H_{1,k}$ and $H_{2,k}$ are given by
\bea H_{1,k} &=& \left( \begin{array}{cc}
0 & J_1 + (J' + \al) e^{-ik} \\
J_1 + (J' + \al) e^{ik} & 0 \end{array} \right), \non \\
H_{2,k} &=& \left( \begin{array}{cc}
0 & J_1 + (J' - \al) e^{-ik} \\
J_1 + (J' - \al) e^{ik} & 0 \end{array} \right). \non \\
\label{ham12} \eea
Looking at the expressions for $H_{1,k}$ and $H_{2,k}$
and using Eq.~\eqref{uk1} to find $H_{F,k}$, it becomes clear
that $H_{F,k}$ will generally be a sum of all the three Pauli
matrices, $\si^x, ~\si^y$ and $\si^z$. It will therefore not
be possible to define a winding number. 

To rectify this situation, 
we consider a different Floquet operator $U'$ which is related to
the earlier operator $U$ by a shift in time by $T/4$, namely,
\beq U' ~=~ e^{-i H_1 T/4} ~e^{-i H_2 T/2}~ e^{-i H_1 T/4}.
\label{u3} \eeq
We then have
\beq U'_k ~=~ e^{-i H_{1,k} T/4} ~e^{-i H_{2,k} T/2} ~
e^{-i H_{1,k} T/4} ~=~ e^{-i H'_{F,k} T}. \label{uk2} \eeq
Since $H_{1,k}$ and $H_{2,k}$ only contain $\si^x$ and $\si^y$,
both of which anticommute with $\si^z$, Eq.~\eqref{uk2} implies
that
\beq (U'_k)^{-1} ~=~ \si^z ~U'_k ~\si^z. \label{iden} \eeq
This identity implies that $U'_k$ must be the exponential of
a matrix consisting of only $\si^x$ and $\si^y$, and hence 
the Floquet Hamiltonian 
\beq H'_{F,k} ~=~ a'_{k,x} \si^x + a'_{k,y} \si^y \label{hfk}
\eeq
contains only two
Pauli matrices. We can therefore define a winding number for $H'_{F,k}$~\cite{tong13,thakur13}.

Before proceeding further, we must discuss an ambiguity in obtaining $H'_{F,k}$
from $U'_k$ through Eq.~\eqref{uk2}. In general, an
$SU(2)$ matrix $U'_k$ can be written as
$e^{i \al_k {\hat n}_k \cdot {\vec \si}}$, where 
${\hat n}_k$ is a unit vector), and its eigenvalues are then given by $e^{\pm i \al_k}$.
The eigenvalues do not change if we shift $\al_k \to 
\al_k + 2 \pi n$ or flip $\al_k \to - \al_k$. We can 
therefore restrict $\al_k$ to lie in the range $[0, \pi]$.
As $k$ varies from 0 to $2 \pi$, the eigenvalues will not
be degenerate if $U'_k$ is not equal to $\pm I$ at any 
value of $k$
(here $I$ denotes the $2 \times 2$ identity matrix). Only
then would it be possible to define a winding number for 
$H'_{F,k}$. (This is the analog of the statement for the
time-independent SSH model that a winding number can be 
defined only if the upper and lower bulk bands do not touch 
each other at any $k$). Hence, assuming that the eigenvalues
of $U'_k$ are not equal to $\pm 1$ for any $k$, we can assume
that $\al_k$ satisfies $0 < \al_k < \pi$ for all $k$.

Next, we note that for $k=0$ and $\pi$, Eqs.~\eqref{ham12}
and \eqref{uk2} imply that $U'_k$ can be written as
\bea U'_0 &=& e^{-i (J_1 + J') T \si^x}, \non \\
U'_\pi &=& e^{-i (J_1 - J') T \si^x}. \label{uk0pi} \eea
If we hold $J_1$ and $J'$ fixed and vary $\om = 2\pi /T$, we 
see that $U'_0$ has degenerate eigenvalues whenever 
$J_1 + J' = p \om /2$, and 
$U'_\pi$ has degenerate eigenvalues whenever $J_1 - J' = p
\om /2$, where $p$ is an integer. At these special values of
$\om$, the winding number becomes ill defined. We
expect the winding number and hence the number of topological
end modes to change abruptly whenever $\om$ crosses one of these 
values. The topological end modes have eigenvalues of $U$ 
equal to $\pm 1$. 

However, as we will see, there can also be 
non-topological modes which are localized at one end of the
system with eigenvalues of $U$ which are {\it not} equal to 
$\pm 1$. These eigenvalues necessarily appear in complex
conjugate pairs, implying that the number of 
non-topological modes at each end must be an even integer.
To prove the complex conjugation property, we note that
there is a unitary transformation $V$ under which $a_j \to
a_j$ and $b_j \to - b_j$ which changes $H \to - H$ in 
Eq.~\eqref{ham1}. (Thus $V$ is a diagonal matrix whose entries 
are equal to $+1$ and $-1$ alternately, implying that 
$V = V^{-1}$). Hence both $U = e^{-i H_2 T/2} e^{- i H_1 T/2}$
and $U = e^{-i H_1 T/4} e^{-i H_2 T/2} e^{-i H_1 T/4}$
satisfy $VU V = U^*$. Then $U \psi = e^{i \ta} \psi$
implies that $U V \psi^* = e^{-i \ta} V \psi^*$.
Hence, if $\psi$ is an eigenstate of $U$ which is localized
at one end with eigenvalue $e^{i \ta}$, $V \psi^*$ will
be an eigenstate of $U$ which is localized at the same end
but has eigenvalue $e^{-i \ta}$. Note that the eigenvalues
of $U$ appearing in complex conjugate pairs is a general
property which is true for both end-localized and bulk eigenstates of $U$.

\subsection{Floquet spectrum and end modes}
\label{sec3a}

We now present our numerical results. First, we discuss the 
Floquet eigenvalues and end modes for a system with open 
boundary conditions.
For our numerical studies, we set $J_1 = 1.1$, $J' = 1$ and
$\al = 0.5$, namely, $J_2 = 1.5$ and $0.5$ for $H_1$ and $H_2$
respectively. As a result, the model is in a topological
phase for $H_1$ and in a on-topological phase for $H_2$. 
We then define the Floquet operator as in Eq.~\eqref{u3}.
We have considered two values of $T$ given by $2 \pi$ and
$\pi/2$ as discussed below.

For $T=2 \pi$, we see in Fig. 2 (a) that there are 
four isolated Floquet eigenvalues $e^{i \ta}$, two of 
which are close to, but not exactly at, $+1$ and the other
two are close to $-1$, for a system with 800 unit cells and 
therefore $L= 1600$ sites). Each of these eigenvalues has
a double degeneracy, corresponding to two modes which are
localized at each end of the system. The probability 
$|\psi_i|^2$ versus the site index $i$
of one the modes localized near the left end
is plotted in Fig. 2 (c). This shows that the mode 
is extremely well localized at the end with a decay
length which is much smaller than the system size. There is negligible hybridization between 
modes localized at opposite ends of the system; hence the 
numerically observed values of the Floquet eigenvalues can be 
assumed to be the same as what they 
would be for an infinitely long system. Therefore, the fact that
the Floquet eigenvalues of these modes are not exactly equal to $\pm 1$
implies that they are not topological end modes. This is
confirmed by looking at the topological invariant which
turns out to be zero in this case; this is discussed in 
Sec.~\ref{sec3b}.

\begin{figure}[H]
\centering
\subfigure[]{
\includegraphics[width=0.8\hsize]{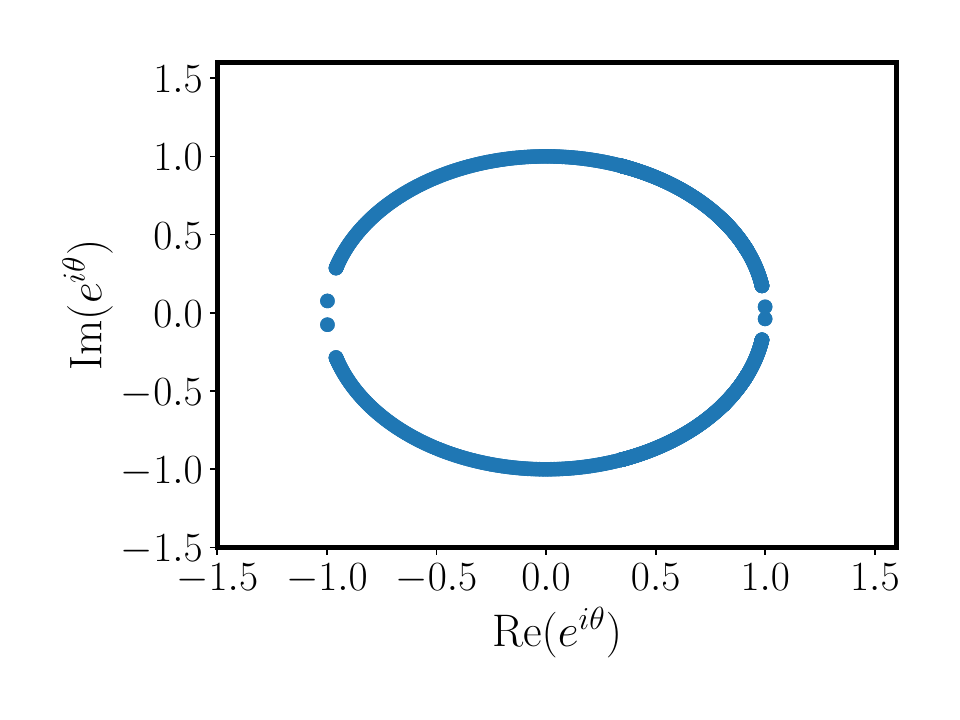}
}
\subfigure[]{
\includegraphics[width=0.8\hsize]{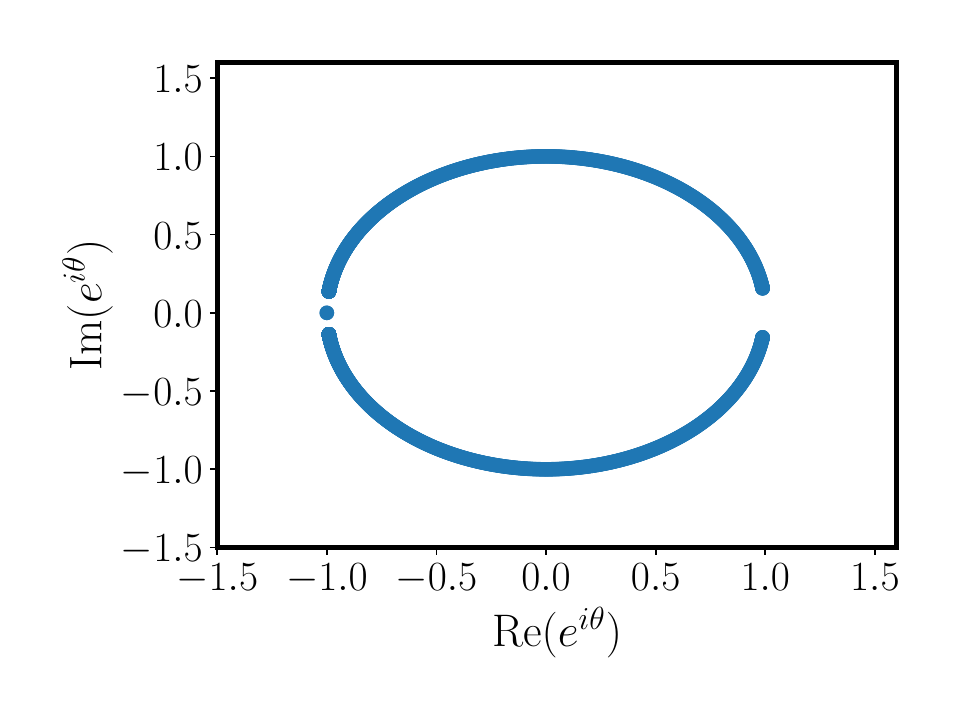}
}
\subfigure[]{
\includegraphics[width=0.8\hsize]{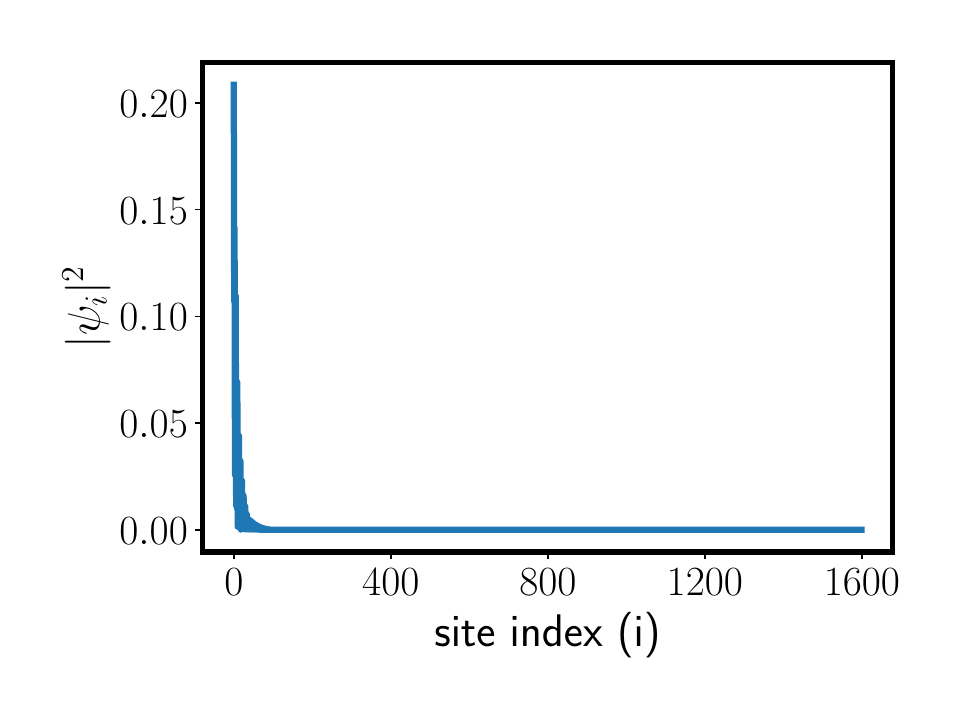}
}
\caption{Plots (a) and (b) show the imaginary part vs real part 
of the eigenvalues $e^{i \ta_n}$ of the Floquet operator for periodic driving with $T = 2\pi$ and $T=\pi/2$ respectively. Plot (c) shows 
the probability $|\psi_i|^2$ versus the site index $i$ of an
eigenvector of the Floquet operator which is localized at 
the left edge of the system, for $T = 2\pi$. (The probability
$|\psi_i|^2$ for the left-localized eigenvector of the 
Floquet operator for $T=\pi/2$ is similar to plot (c) and
is not shown here). The parameters are $J_1 = 1.1$, 
$J'= 1$, $\alpha = 0.5$, and $L = 1600$.}
\label{fig2} \end{figure}

For $T=\pi/2$, Fig. 2 (b) shows that there is
an isolated Floquet eigenvalue
which lies exactly at $e^{i \ta} = -1$, 
for a system with 1600 sites. This eigenvalue
has a double degeneracy, corresponding to one mode
localized at each end of the system. The fact that
the Floquet eigenvalue of these modes is exactly equal 
to $- 1$ implies that it is a topological end mode;
this is confirmed by the topological invariant which
turns out to be $-1$ in this case (see Sec.~\ref{sec3b}).

\subsection{Topological invariant}
\label{sec3b}

\begin{figure}[H]
\centering
\subfigure[]{
\includegraphics[width=0.75\hsize]{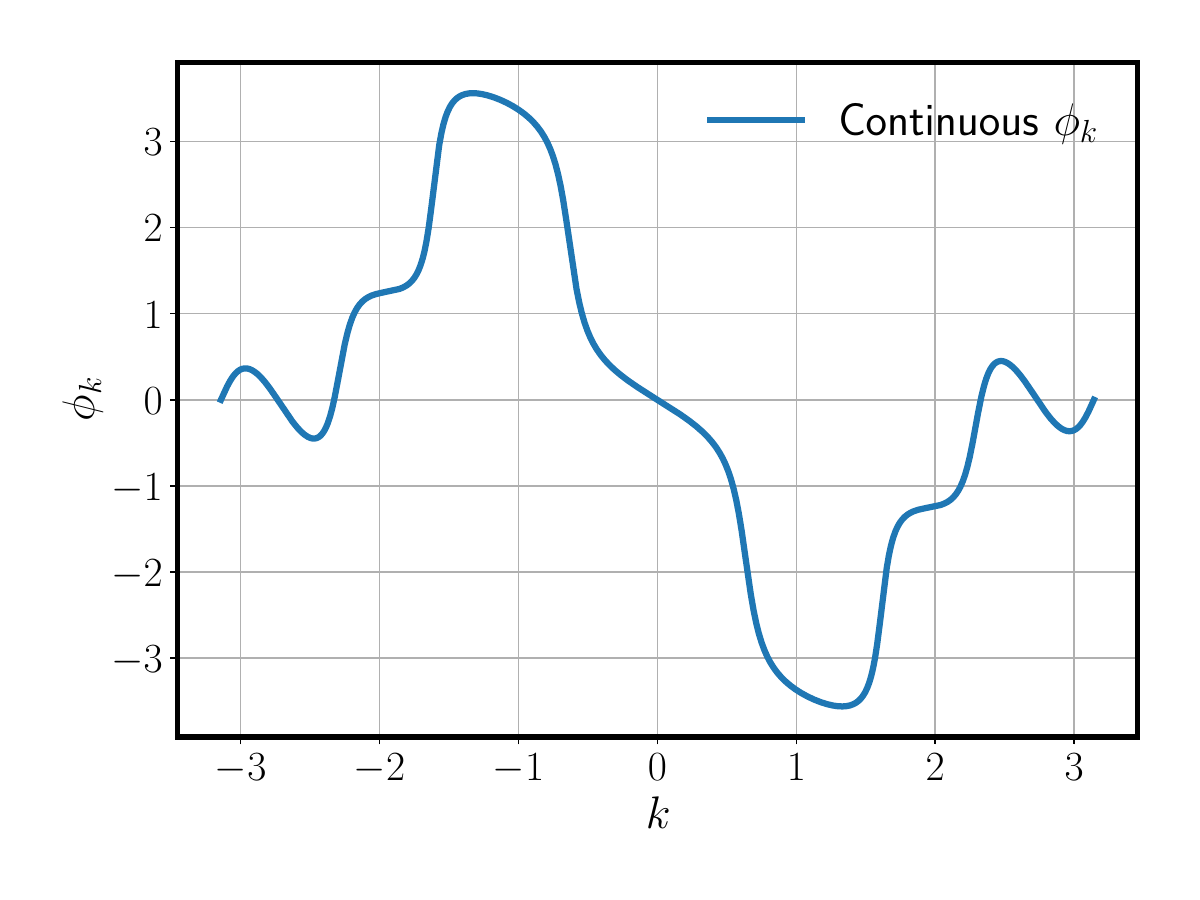}
}\\
\subfigure[]{
\includegraphics[width=0.75\hsize]{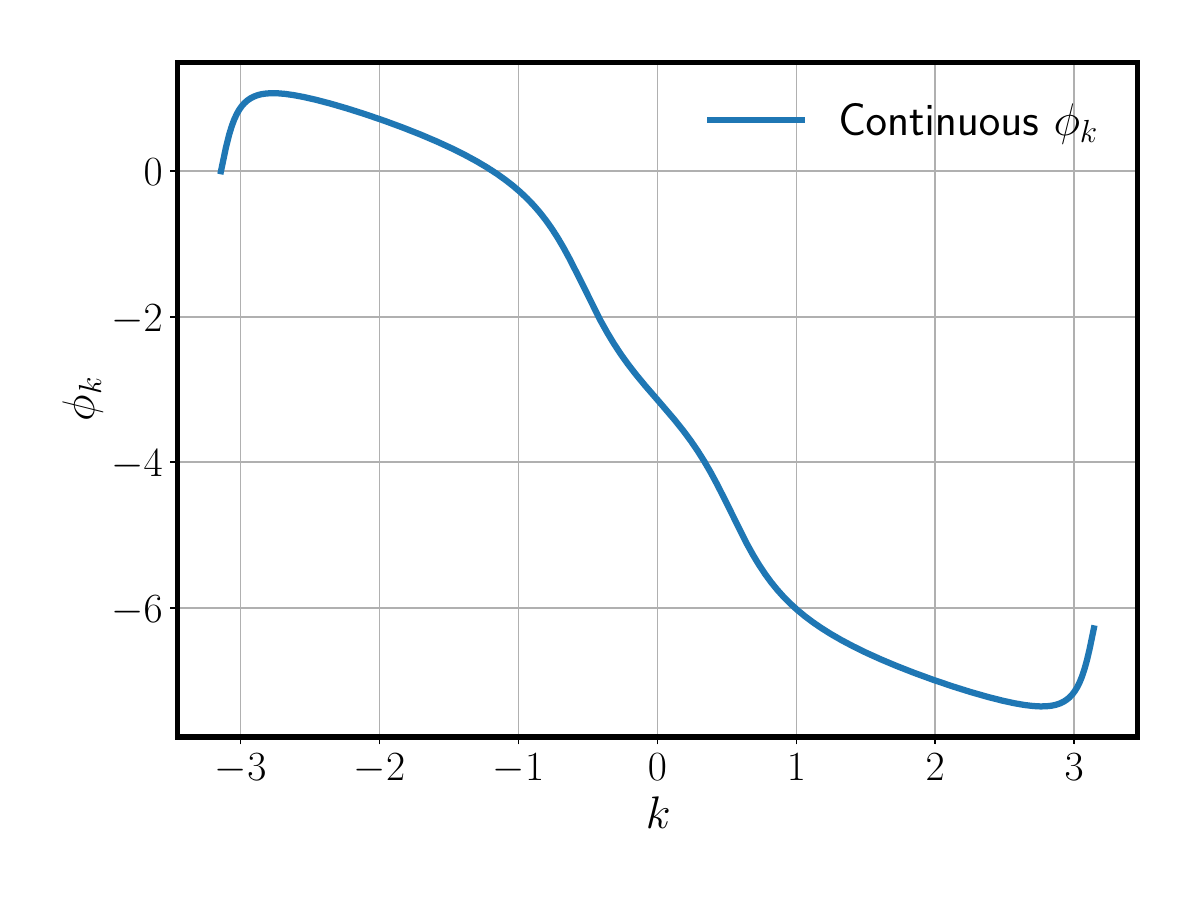}
}
\caption{Plot of $\phi_k$ versus $k$ for periodic driving 
with (a) $T=2 \pi$ and for (b) $T=\pi/2$ showing $\phi_k$ as a function of $k$. 
The parameters used are $J_1 = 1.1$, $J'=1$, $\alpha = 
0.5$, and $L = 1600$. Plot (a) shows that $W=0$,
while plot (b) shows that $W=-1$.}
\label{fig3} \end{figure}

We now calculate the winding number $W$ numerically.
To do this, we numerically calculate the Floquet operator $U'_k$
using the symmetrized driving protocol 
described in Eq.~\eqref{uk2}. Then $U'_k$ involves only
two Pauli matrices as
\beq U'_k ~=~ e^{-i T (a_{k,x} \si^x + a_{k,y} \si^y)}
\eeq
where 
\beq a_k = \sqrt{ a^2_{k,x} ~+~ a^2_{k,y} }, \eeq
and we can assume that $0 < a_k < \pi/T$ as discussed after
Eq.~\eqref{hfk}. The parameters $a_k, ~a_{k,x}$ and
$a_{k,y}$ can be computed from the numerically obtained 
value of
\beq U'_k ~=~ \begin{bmatrix}
U'_{k,11} & U'_{k,12}\\
U'_{k,21} & U'_{k,22}
\end{bmatrix}, \label{uij} \eeq
as
\bea a_k &=& \frac{1}{T} ~\cos^{-1} [U'_{k,11}], \non \\
a_{k,x} &=& \frac{i a_k}{2 \sin (a_k T)} ~(U'_{k,12} + U'_{k,21}), \non \\
a_{k,y} &=& - ~\frac{a_k}{2 \sin (a_k T)} ~(U'_{k,12} - U'_{k,21}). \label{aphik} \eea
If we find $(a_{k,x},a_{k,y})$ for $N$ equally spaced values
of $k$ from $- \pi$ to $\pi$ labeled as 1 to $N$ (assumed
to be sufficiently large), then the winding number can
be found from the discretized form of the second integral in
Eq.~\eqref{wind} as
\beq W = \frac{1}{ 2\pi} \sum_{n=1}^N \frac{a_{k,x} (n) a_{k,y}(n+1) - a_{k,y}(n) a_{k,x}(n+1)}{a^2_k (n)}.
\eeq
Alternatively, we can calculate $\phi_k = \tan^{-1} (a_{k,y}/a_{k,x})$, interpolate $\phi_k$ to obtain a 
continuous function of $k$, and then compute the winding number $W$ as shown in the first integral in Eq.~\eqref{wind}.
In fact, we can just look at
plots of $\phi_k$ versus $k$ as shown in Fig.~\ref{fig3}, and read off the winding number
as $W= (\phi_{\pi} - \phi_{-\pi}) /(2 \pi)$. For $T=2 \pi$,
we find that $W=0$ implying that there are no topological
end modes, while for $T= \pi/2$,
we find $W= - 1$ implying that there must one topological
mode at each end of the system. This is in agreement with the 
results presented in Sec.~\ref{sec3a}.

\subsection{Loschmidt amplitude}
\label{sec3c}

The Loschmidt echo ($LE$) or return probability is a well-studied and important quantity to understand the long-time behavior of a state. It characterizes how much memory of the initial state
is retained at long times. If $| \phi (0) \ra$ denotes the
initial state and the state obtained after evolving for a 
time $t$ is denoted by \textcolor{blue}{$| \phi (t) \ra$}, the $LE$ is defined as $|\langle \phi(t)| \phi (0) \rangle |^2$. In this 
section, however, we will study the quantity $\langle \phi(t)| \phi (0) \rangle$ called the
Loschmidt amplitude ($LA$). We will see below why the
$LA$ has a simpler structure than the $LE$.

\begin{figure}[H]
\centering
\subfigure[]{
\hspace*{-1.2cm} \includegraphics[width=0.85\hsize]{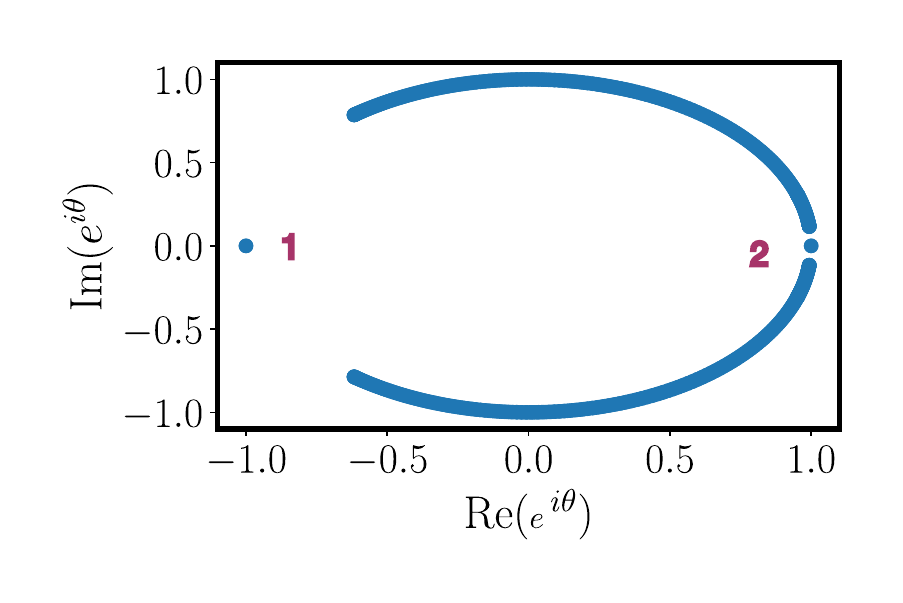}
}\\
\subfigure[]{
\hspace*{-1.2cm} \includegraphics[width=0.85\hsize]{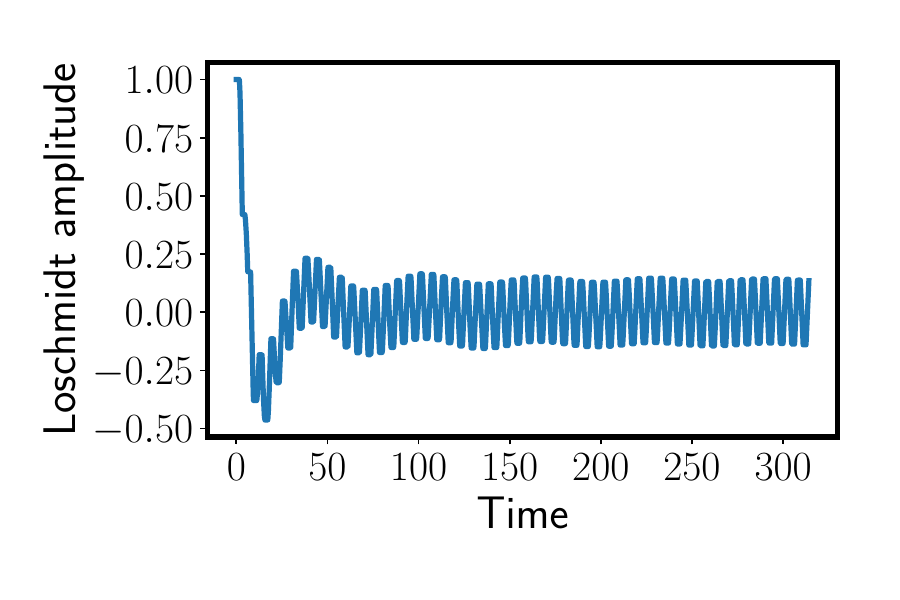}
}\\
\subfigure[]{
\hspace*{-1.2cm} \includegraphics[width=0.85\hsize]{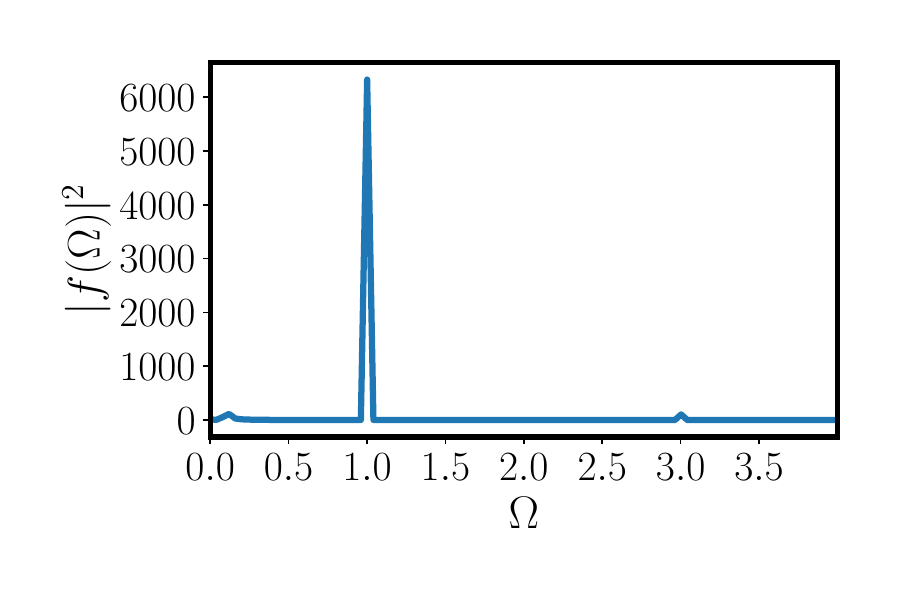}
}
\caption{Plot (a) shows the imaginary part versus real part of the eigenvalues of the Floquet operator for periodic driving. The points marked
1 and 2 show the Floquet eigenvalues of end modes
at $e^{i \ta} = -1$ and $e^{i \ta} = 1$ respectively.
Plot (b) shows the 
Loschmidt amplitude for an initial state which is taken to be
the left-localized end mode of $H_1$. Plot (c) shows the Fourier transform of the 
Loschmidt amplitude. We see a sharp peak at $\Omega = 1$.
The parameters used for these plots are $J_1 =1.1$, $J' = 1$,
$\al = 0.5$, $T = \pi$, and $L=800$.} \label{fig4} \end{figure}

For a system driven with a time period $T$, it is convenient
to study the $LA$ at stroboscopic times given by $t=n T$, where
$n$ is an integer; then $| \phi (nT) \rangle = U^n | \phi (0) \rangle$, where $U = U_2 U_1$ is the Floquet operator.
For our study we have chosen $J_1 = 1.1, ~J_2 = 1 \pm 0.5, ~T = \pi$ and $L=800$. (This system size ensures that there is
negligible hybridization between the left-localized and
right-localized end modes). The Floquet eigenvalues $e^{i \theta_m} = 
e^{-i E_m T}$ are shown in Fig.~\ref{fig4} (a). The two points
1 and 2 marked in the plot respectively denote 
an end mode with quasienergy $E=\pi/T$ and an end mode
with quasienergy $E=0$; we will denote these Floquet eigenstates
as $| \psi_1 \ra$ and $| \psi_2 \ra$ respectively.
We will numerically calculate the $LA$
for an initial state $| \phi (0) \rangle$ which is
chosen to be the left-localized end mode of $H_1$ 
(where $U_1 = e^{-i H_1 T/2}$). 
The $LA$ is shown in Fig.~\ref{fig4} (b); note that it is
real for reasons which are explained below.
We see regular oscillations in the range of about
$70 < t < 320$. To determine the oscillation frequency, we 
do a Fourier transform to go from $LA (t)$ to $f(\Omega)$. 
The modulus squared of this is shown in Fig.~\ref{fig4} (c). We see a sharp peak in $|f(\Omega)|^2$ at $\Omega = 1$.

The presence of the peak can be
understood from the spectrum of Floquet eigenvalues. If the initial state 
$| \phi (0) \rangle$ is written in terms of the Floquet
eigenvectors $| u_m \ra$ (with Floquet eigenvalue $e^{- i E_m T}$, where $E_m$ denotes the Floquet quasienergy) as
\beq | \phi (0) \rangle ~=~ \sum_m ~c_m | u_m \rangle, \label{psirand} \eeq
then the $LA$ at time $t=nT$ is given by 
\bea LA (t) &=& \bra{\phi (0)} U^n \ket{\phi (0)}
\non \\
&=& \sum_m |c_m|^2 e^{- i E_m nT}. \label{LA} \eea
(We note that the $LA$ is real in our case, since the initial state has a real wave function, the Floquet eigenvalues
come in complex conjugate pairs $e^{\pm i E_m T}$, and the
overlaps $|c_m|^2$ of the corresponding eigenvectors $\psi_\pm$ 
with the initial state are equal in magnitude since
$\psi_+ = V \psi_-^*$). Equation~\eqref{LA} shows that
the Fourier transform of the $LA$ directly gives the
spectrum of Floquet quasienergies for those Floquet eigenvectors
which have a significant overlap with the initial state.
The oscillations in the modulus squared of the Fourier transform 
of $LA$, $|f (\Omega)|^2$, will be dominated by terms for which
$|c_m|^2$ is large, i.e., by states $m$ which have a large
overlap with the initial state. We now note that $|\phi (0) \rangle$, 
the left-localized end mode of $H_1$, has a large overlap with 
the left-localized end modes of $U= U_2 U_1$. Since there
are left-localized end modes at Floquet quasienergies equal to $0$ and $\pi/T$, we expect the Fourier transform of the
$LA$ to have peaks at those values of $\Omega$. In 
Fig.~\ref{fig4} (c) we see a large peak at $\Omega = 1$ but 
there is no peak at $\Omega = 0$, suggesting that the initial
state has a significant overlap with the end mode $| \psi_1 \ra$
at $E_1 = \pi/T = 1$ and very little overlap with the end mode
$| \psi_2 \ra$ at $E_2 = 0$. Indeed, we find that $| \la \psi_1 | \phi (0)\ra |^2 = 0.139$ and $| \la \psi_2 | \phi (0) \ra |^2 = 
0.000$ to three decimal places.

Before ending this section, we remark that the $LA$ is a simpler quantity to study compared to 
the $LE$ which is defined as
\bea LE (t) &=& |\bra{\phi (0)} U^n \ket{\phi (0)}|^2 
\non \\
&=& \sum_{m,p} |c_m|^2 |c_p|^2 e^{i (E_p - E_m)nT}. \label{LE}
\eea
The expression in Eq.~\eqref{LE} contains an exponential
$e^{i (E_p - E_m)nT}$ for every {\it pair} of states $m$ and $p$,
and therefore gets a large contribution only if both the
states $m$ and $p$ have large overlaps with the
initial state. Hence the $LE$ and its Fourier transform have
a significantly more complicated structure than the $LA$.
However, in the following sections, we will find it more convenient
to study the $LE$.

\section{Fibonacci driving with two unitaries}
\label{sec4}

\begin{figure}[H]
\centering
\subfigure[]{
\hspace*{-1cm} \includegraphics[scale=0.45]{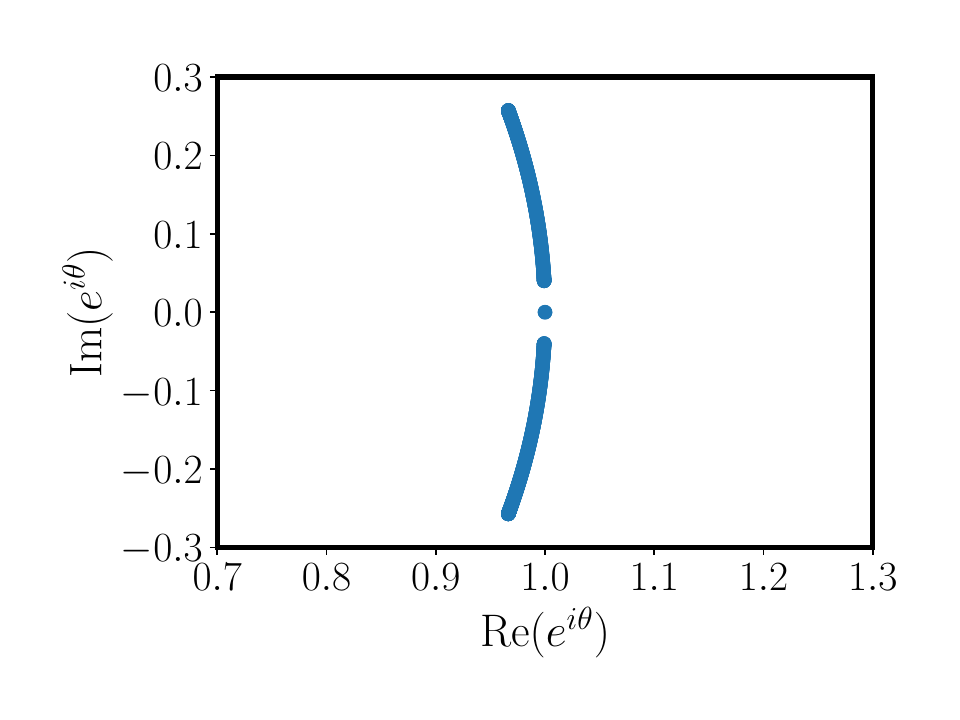}
}
\hfill
\subfigure[]{
\hspace*{-1cm} \includegraphics[scale=0.45]{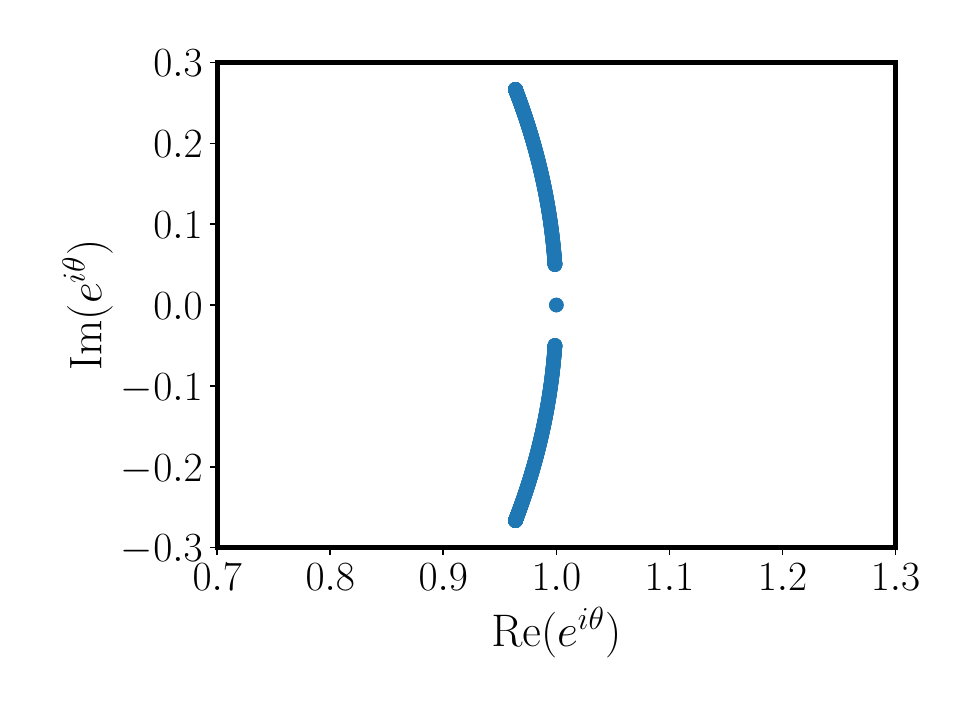}
}
\caption{Plots of the real versus imaginary parts of the Floquet eigenvalues of (a) $U_1$ and (b) $U_2$. We have taken
$J_1 = 1.1$ and $T=0.1$ in both cases, while $J_2 = 1.5$ 
and $J_2 = 1.5 + \ep$ for $U_1$ and $U_2$ respectively,
with $\ep = 0.1$. Both $U_1$ and $U_2$ have one mode
localized et each end of an open system, with
eigenvalue equal to 1 as shown.} \label{fig5} \end{figure}

\begin{figure}[H]
\centering
\hspace*{-1cm} \includegraphics[scale=0.45]{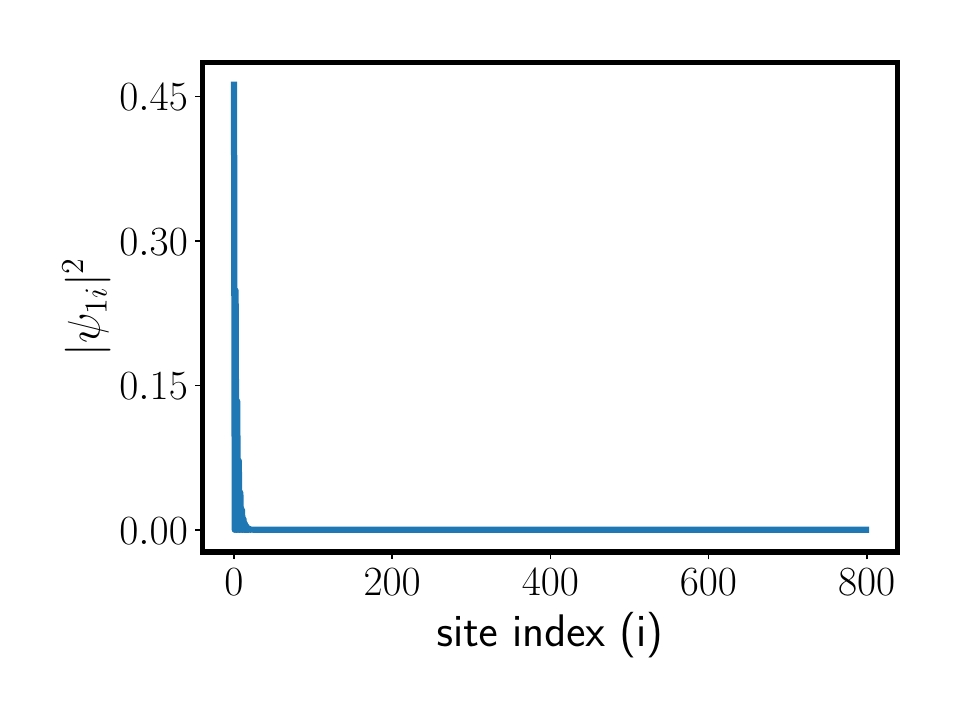}
\caption{Plot of $|\psi_i|^2$ for the left-localized end modes of $U_1$.
The system parameters are 
the same as in Fig.~\ref{fig5}.} \label{fig6} \end{figure}

In this section, we will study the dynamics of end modes
when the system is driven by a quasiperiodic Fibonacci sequence of two
unitaries $U_1 = e^{-i H_1 T}$ and $U_2 = e^{-i H_2 T}$ which differ slightly from each other. We will take
$H_1$ and $H_2$ to be Hamiltonians for the SSH model,
with the parameters being $J_1 = 1.1$, $J_2 = 1.5$
for $H_1$ and $J_1 = 1.1$, $J_2 = 1.5 + \ep$ for $H_2$,
where $\ep$ is a small number. (The values of $\ep$ and the
time $T$ will be specified below). For these values
of the parameters, 
both $H_1$ and $H_2$ have exactly one 
end mode at each end of an open system; we will denote
the mode localized at the left end as
$| \psi_1 \ra$ and $| \psi_2 \ra$ for $H_1$ and $H_2$ 
respectively. Since these modes are eigenstates of $H_1$
and $H_2$ with zero eigenvalue, they are eigenstates of 
$U_1$ and $U_2$
with eigenvalue 1, and they are separated from the 
eigenvalues of the bulk states by a gap. This is shown in
Fig.~\ref{fig5} where we have taken $\ep = 0.1$ and $T= 
0.1$. The plot of the probabilities $|\psi_i|^2$
of the eigenstate of $U_1$ which is localized near the
left end is shown in Fig.~\ref{fig6}. The plot for the
left-localized end mode
of $U_2$ is very similar and is not shown.

We will examine what happens if we begin with an
initial state $| \psi_1 \ra$ which is the left-localized
state of $U_1$, and then $U_1$ and 
$U_2$ act on it following a Fibonacci sequence. 
To generate such a sequence we use the following rule~\cite{kraus12}. 
We define 
\beq b ~=~ \frac{1}{2} ~(\sqrt{5}-1) ~\simeq~ 0.618, 
\label{gold} \eeq
and a function
\beq f_j = \cos(2 \pi b j + \pi b) - \cos (\pi b), \label{fj} \eeq
where $j=1,2,3,\cdots$. We first note that $f_j = 0$ does not
occur for any integer value of
$j$ in Eq.~\eqref{fj} because $b$ is irrational.
Next, we define
\bea U(f_j) &=& U_1 ~~{\rm if}~~ f_j ~>~ 0, \non \\
&=& U_2 ~~{\rm if}~~ f_j ~<~ 0. \label{ufj} \eea
We then act on $| \psi_1 \ra$ with $U(f_1), ~U(f_2),
~U(f_3), \cdots$, in that order. Namely, the state
obtained after $j$ drives is given by
$U(f_j) U(f_{j-1}) \cdots U(f_2) U(f_1) | \psi_1 \ra$.

The above rule for defining a Fibonacci sequence
is exactly equivalent to one in which we act with 
sequences of unitaries which are
recursively generated as follows: we define $V_1 = U_1$,
$V_2 = U_2 U_1$, and then $V_j = V_{j-2} V_{j-1}$ for $j \ge 3$.
The first few sequences are given by
\bea V_1 &=& U_1, ~~V_2 = U_2 U_1, ~~V_3 = U_1 U_2 U_1,
\non \\
V_4 &=& U_2 U_1 U_1 U_2 U_1, ~~V_5 = U_1 U_2 U_1 U_2 U_1 U_1 U_2 U_1, \label{fs}
\eea
and so on. The number of unitaries appearing in $V_j$ is
equal to the Fibonacci number $F_j$ which satisfies $F_1 = 1$,
$F_2 = 2$, and $F_j = F_{j-2} + F_{j-1}$ for $j \ge 3$.
As $j \to \infty$, the Fibonacci numbers grow exponentially as a constant times $b^j$. 

Depending on the quantity and timescale of interest, we will work with either the
sequences generated by $U (f_j)$ in Eq.~\eqref{ufj} 
whose lengths grow linearly with $j$, or the sequences generated by $V_j$ given in Eq.~\eqref{fs}
whose lengths grow exponentially with $j$.

\subsection{Numerical results}
\label{sec4a}

\begin{figure}[H]
\centering
\includegraphics[width=0.9\linewidth]{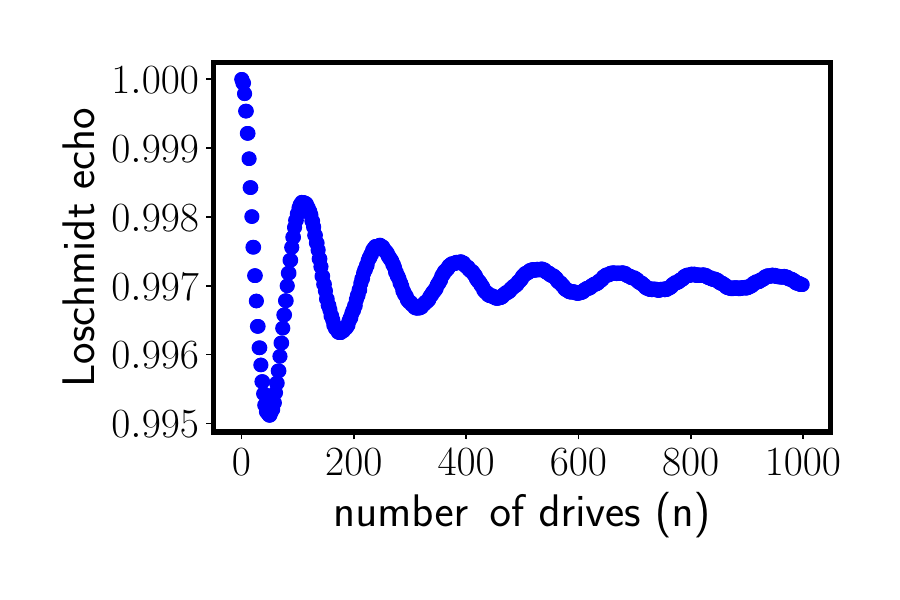}
\caption{Plot showing the initial oscillations of $LE$ for 
Fibonacci driving with $J_1 = 1.1, ~J_2 = 1.5, ~\ep = 0.1$, 
$T = 0.1$, and $L=800$.} \label{fig7} \end{figure}

\begin{figure}[H]
\centering
\subfigure[]{
\includegraphics[scale=0.47]{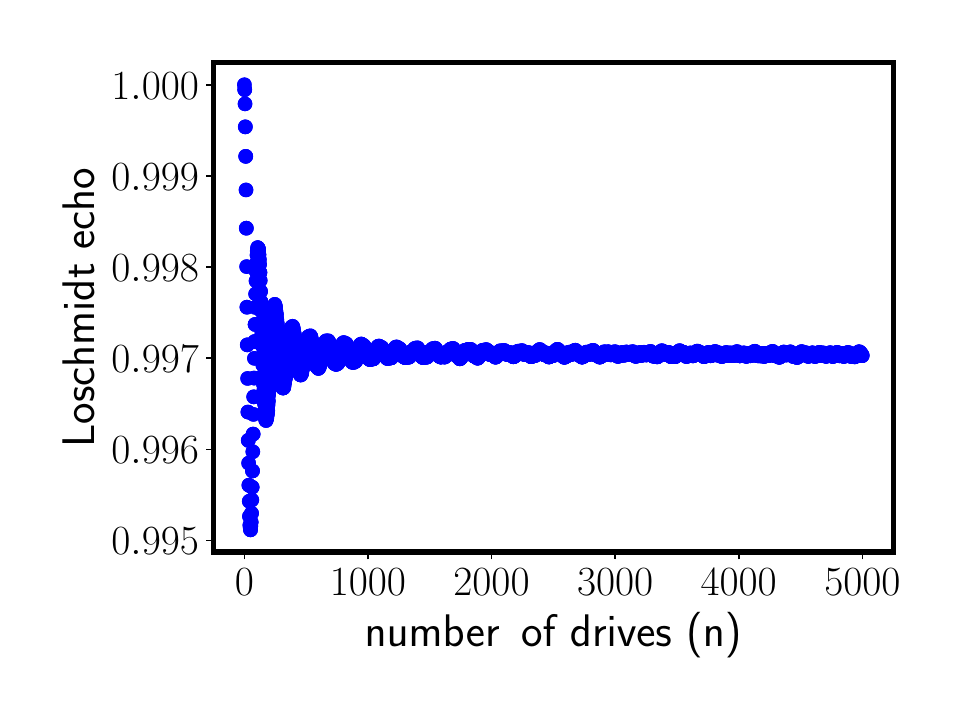}
}
\subfigure[]{
\includegraphics[scale=0.47]{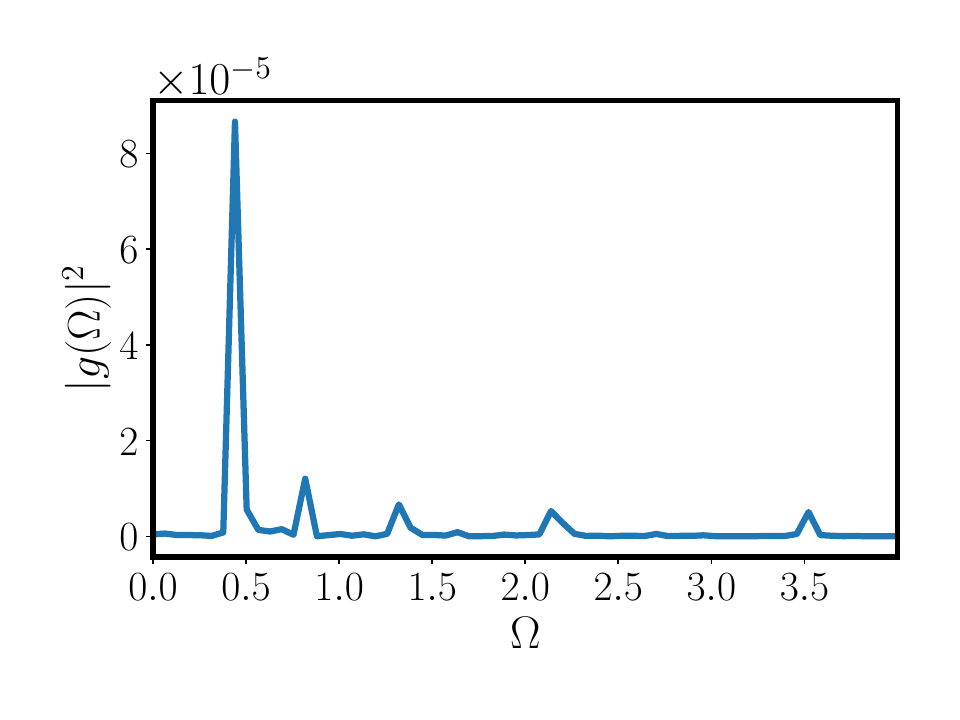}
}
\caption{Plot (a) shows the $LE$ for Fibonacci driving 
of the end mode for $J_1 = 1.1, ~J_2 = 1.5,~ \ep = 0.1$, 
$T=0.1$, and $L=800$. Plot (b) shows the modulus squared of the Fourier transform, $|g(\Omega)|^2$, of the $LE$. 
We see a peak around $\Omega = 0.5$.} 
\label{fig8} \end{figure}

We now present our numerical results. We begin by 
using Eq.~\eqref{ufj} to act upon $| \psi_1 \ra$ by a 
Fibonacci sequence of length $n$. This gives us a state $| \psi_1 (t) \ra$, where $t=nT$. Upon computing the Loschmidt echo 
defined as $LE = |\bra{\psi_1(t)}\psi_1\rangle|^2$, we
find that it stays close to 1 and saturates at some value when $n$ becomes large,
if $\ep$ and $T$ are both small. This is shown in 
Fig.~\ref{fig7} for $J_1 = 1.1, ~J_2 = 1.5, ~\ep = 0.1$, and $T = 0.1$; we see that the $LE$ saturates, with some
oscillations, at about
$0.997$ when $n$ reaches about $1000$. The oscillations
continue to be visible up to $n = 5000$ as shown in 
Fig.~\ref{fig8} (a).

We now examine in more detail the oscillations in
the LE shown in Fig.~\ref{fig8} (a). To see if these 
oscillations are related to the Floquet
energy spectrum of either $U_1$ or $U_2$, we do a Fourier
transform of the LE between drive numbers $n=n_i$ and $n_f$, corresponding to times $t_i = n_i T$ and
$t_f = n_f T$ respectively; this spans a total time interval equal to $t_f - t_i = N_p T$, where
$N_p = n_f - n_i$. (For our numerical analysis 
we have taken $n_i = 300$, which is large enough to
avoid the initial transient behavior, and $n_f = 1300$). 
The Fourier transform takes us from the variable
$LE (n) = LE (t = n T)$ to $g (k) = g(\Omega = 2 \pi
k /(N_p T))$, where both $n$ and $k$ can take $N_p -1$ 
possible values. The Fourier transform is then defined in a standard way as
\bea g(\Omega ) &=& \sum_{n = n_i+1}^{n_f} ~e^{-i 
2\pi n k /N_p} ~LE (n) \non \\
&=& \sum_{n=n_i+1}^{n_f} ~e^{-i \Omega t} ~LE (t). \eea

Figure~\ref{fig8} (b) shows the modulus squared of the Fourier transform, $| g(\Omega)|^2$ of the $LE$ shown in
Fig.~\ref{fig8} (a). We see a very prominent peak close
to $\Omega = 0.5$, and several smaller peaks at larger
values of $\Omega$. 
Given the expression for the $LE$ in Eq.~\eqref{LE},
we see that $\Omega 
\simeq 0.5$ corresponds to the Floquet quasienergy gap,
$\Delta \Omega = E_p - E_m$, between the
end mode lying at zero quasienergy and the nearest
bulk mode as we can see in Fig.~\ref{fig5} for either $U_1$
or $U_2$. Namely, the Floquet eigenvalue gap $\Delta 
\theta \simeq 0.05$ in Fig.~\ref{fig5} is 
approximately equal to $T=0.1$ times $\Omega = 0.5$ which 
is the location of the first peak in Fig.~\ref{fig8} (b).

\begin{figure}[htb]
\centering
\subfigure[]{
\includegraphics[scale=0.48]{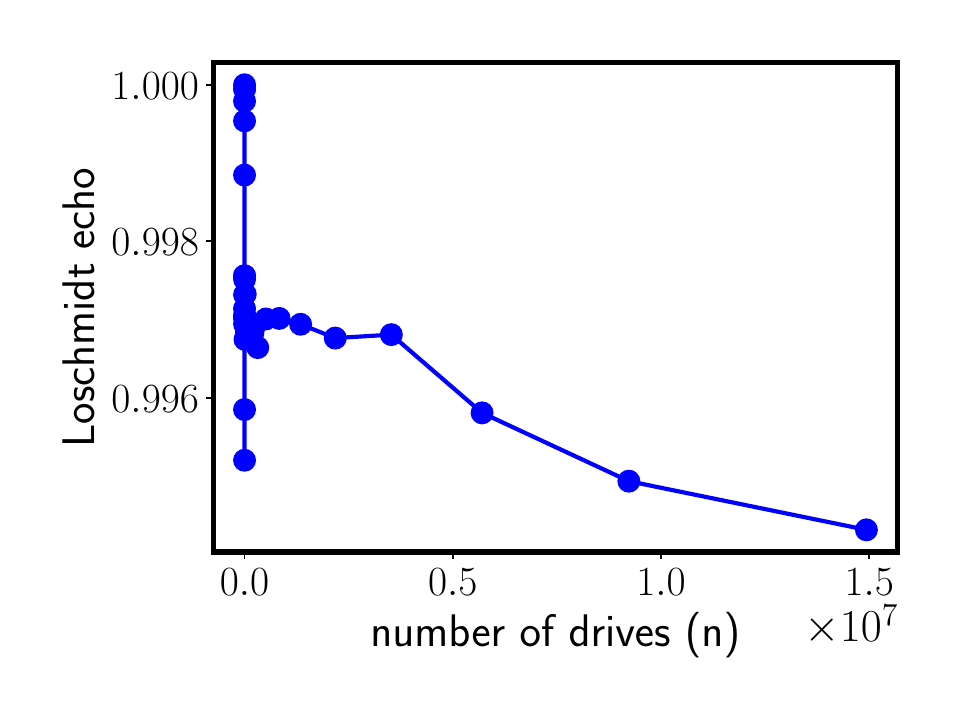}
}
\subfigure[]{
\includegraphics[scale=0.48]{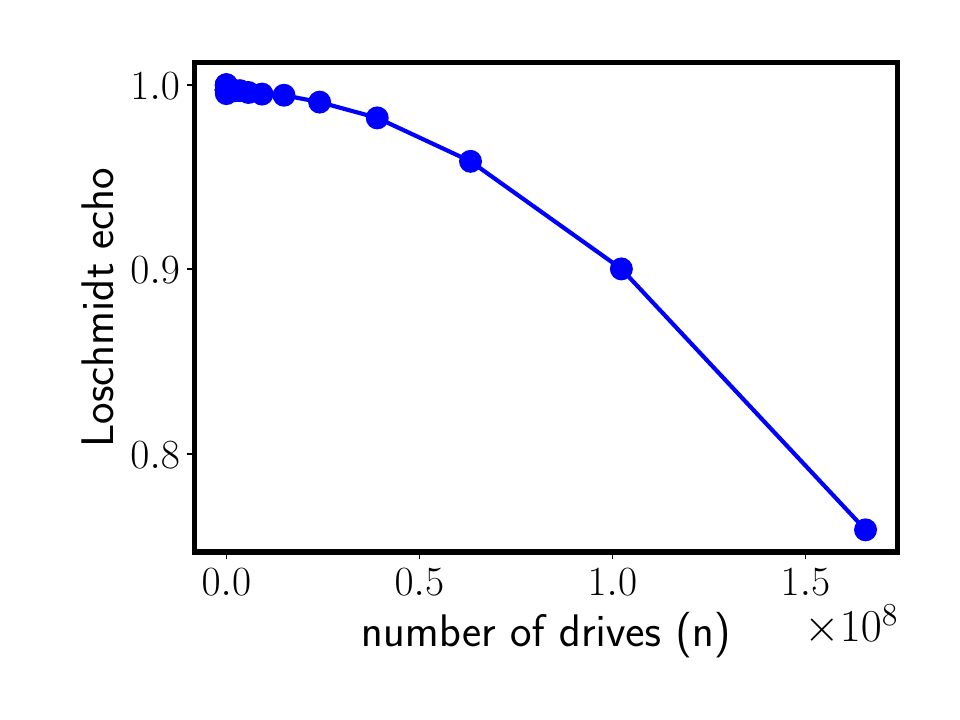}
}
\caption{Plots of the $LE$ for Fibonacci driving after acting
with an exponentially large number of drives $n$ on the initial 
state $|\psi_1 \ra$, going up to about $1.5 \times 10^7$ in
(a) and $1.6 \times 10^8$ in (b). The value of $LE$ remains 
close to $1$ even when $n$ is as large as $10^7$. However, 
the $LE$ starts deviating appreciably from 1 when $n$
becomes much larger than $10^7$. The system
parameters are the same as in Fig.~\ref{fig8}.}
\label{fig9} \end{figure}

When $n$ becomes extremely
large, of the order of $10^7$, the $LE$ starts deviating
from the saturation value seen in Fig.~\ref{fig8} (a). 
This is shown in Fig.~\ref{fig9} which is generated by
using Eq.~\eqref{fs} to generate exponentially long
sequences. We note that all these results hold only if
$T$ is small. If $T$ is of order 1, we find numerically
that the $LE$ deviates rapidly from 1 as the number of
drives increases.

\subsection{Dependence of the distance between $U_1$ and $U_2$ on the parameters $\ep$ and $T$}
\label{sec4b}

\begin{figure}[htb]
\centering
\subfigure[]{
\includegraphics[scale=0.48]{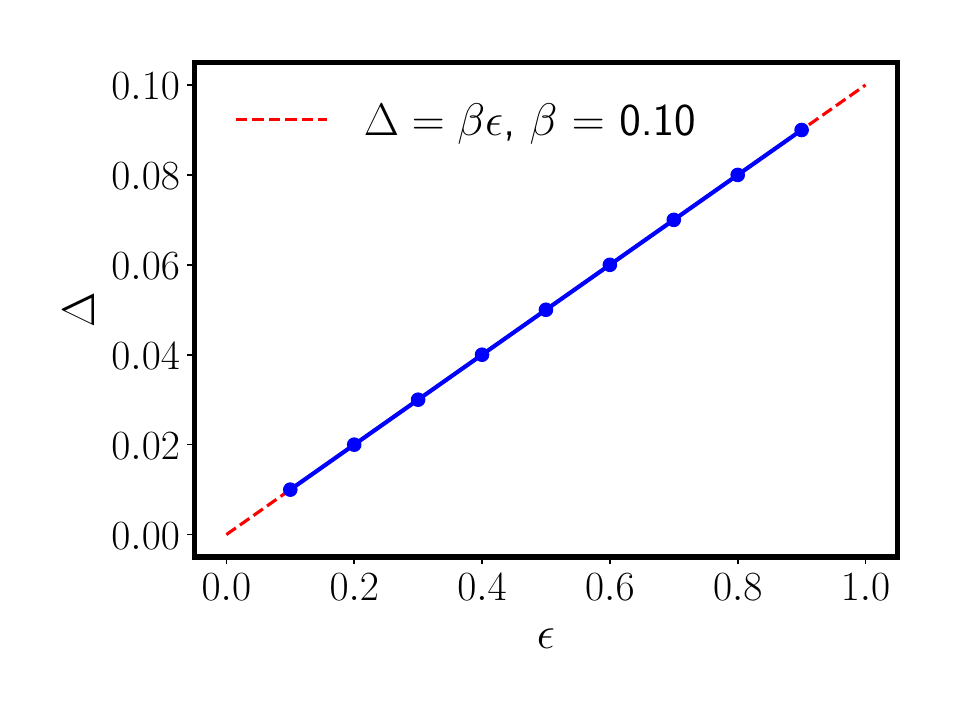}
}
\hfill
\subfigure[]{
\includegraphics[scale=0.48]{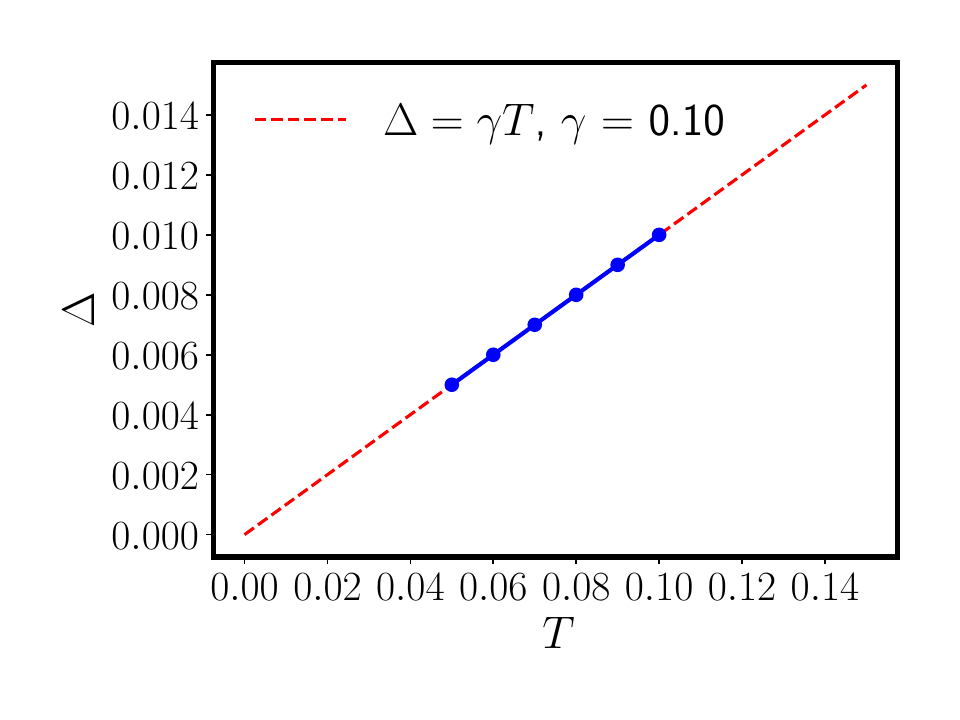}
}
\caption{(a) Plot of $\Delta$ versus $\ep$ for fixed 
$T = 0.1$. (b) Plot of $\Delta$ versus $T$ for fixed 
$\ep = 0.1$. In both cases we have taken $J_1 = 1.1$ 
and $J_2 = 1.5$.} \label{fig10} \end{figure}

To understand why the $LE$ stays close to 1 for a 
very long number of drives when $\ep$ and $T$ are
small, it is useful to understand how close the
unitaries $U_1 = e^{-i H_1 T}$ and $U_2 = e^{-i H_2 T}$ are to each other. Clearly, $U_1$ is identically
equal to $U_2$ if either $\ep =0$ or $T = 0$ (we recall
that $H_1$ and $H_2$ differ by a term of order $\ep$).
This leads us to define the distance between $U_1$ and $U_2$ as
follows. We first define a matrix $M = U_1 U_2^{-1} 
- I$, where $I$ denotes the identity matrix. Then
$M$ will be a matrix of order $\ep T$ if
$\ep$ and $T$ are both small. The
distance between $U_1$ and $U_2$ is then defined as
the spectral norm, $ \Delta = \sqrt{max ( singular\,\, values (M^\dagger M))}$. Then $\Delta$ will be of order 
$\ep T$ if $\ep$ and $T$ are small. Hence, $\Delta$ is proportional to $T$ if $\ep$
is held fixed and proportional to $\ep$ if $T$ is held fixed. 
This is shown in Tables 1 and 2 and is illustrated in Fig.~\ref{fig10} for a system
with $J_1 = 1.1$ and $J_2 = 1.5$.

\begin{widetext}
\begin{center}
\begin{tabular}{|c|c|c|c|c|c|c|c|c|c|}
\hline
 $\ep$ & 0.1 & 0.2 & 0.3 & 0.4 & 0.5 & 0.6 & 0.7 & 0.8 & 0.9\\
 \hline
 $\Delta$ & 0.0099& 0.0199 & 0.0299 & 0.0399 & 0.0499 & 0.0599 & 0.0699 & 0.0799 & 0.0899 \\
 \hline
\end{tabular} \\
\vspace*{.2cm}
Table 1: $\Delta$ versus $\ep$,
for $J_1 = 1.1$, $J_2=1.5$, and $T=0.1$.
\end{center}
\vspace*{.2cm}

\begin{center}
\begin{tabular}{|c|c|c|c|c|c|c|c|c|c|}
\hline
 $T$ & 0.05 & 0.06 & 0.07 & 0.08 & 0.09 & 0.1 \\
 \hline
 $\Delta$ & 0.0049 & 0.0059 & 0.0069 & 0.0079 & 0.0089 & 0.0099 \\
 \hline
\end{tabular} \\
\vspace*{.2cm}
Table 2: $\Delta$ versus $T$,
for $J_1 = 1.1$, $J_2=1.5$, and $\ep =0.1$.
\end{center}
\vspace*{.2cm}

\begin{center}
\begin{tabular}{|c|c|c|c|c|c|c|c|c|c|c|}
\hline
 $\ep$ & 0.01 & 0.02 & 0.03 & 0.04 & 0.05 & 0.06 & 0.07 & 0.08 & 0.09 & 0.1\\
 \hline
 $LE_{sat}$ & 0.999967 & 0.99987 & 0.99971 & 0.999497 & 0.999216 & 0.99888 & 0.998493 & 0.99806 & 0.99756 & 0.99703 \\
 \hline
 d & 0.000033 & 0.00013 & 0.00029 & 0.000503 & 0.000784 & 0.00112 & 0.001507 & 0.00194 & 0.00244 & 0.00297 \\
 \hline
 \end{tabular}
\end{center}
Table 3: $LE_{sat}$ (the saturation value of 
$LE$) and $d= 1 - LE_{sat}$ versus $\ep$,
for $J_1 = 1.1$, $J_2 = 1.5$, and $T=0.1$.
\end{widetext}

\subsection{Scaling of the saturation value of $LE$ with
$\ep$}
\label{sec4c}

\begin{figure}[htb]
\centering
\includegraphics[width=0.9\linewidth]{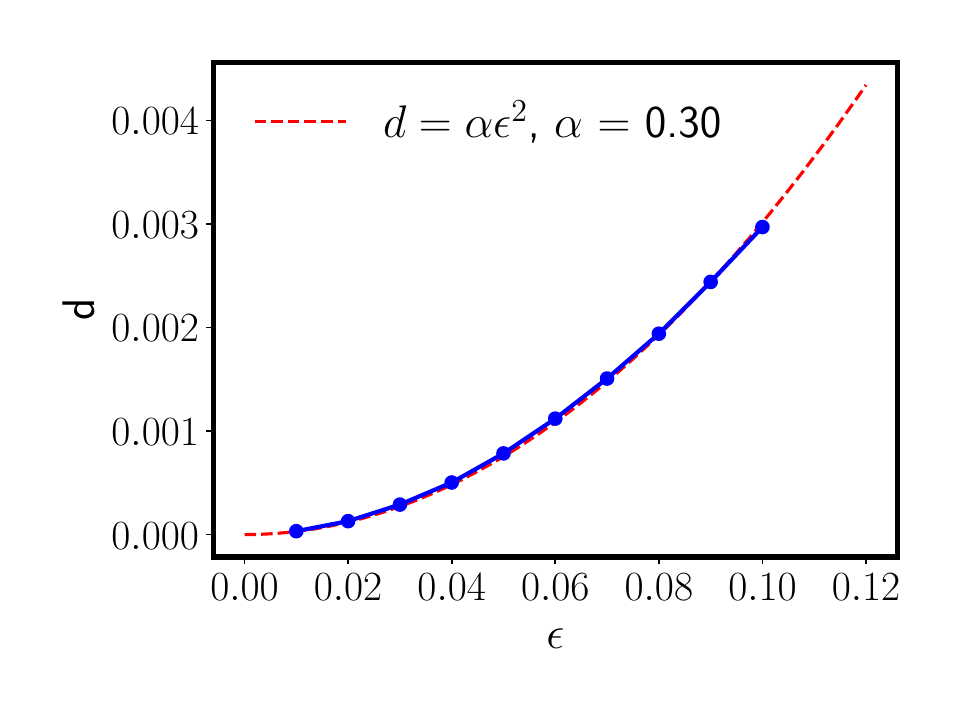}
\caption{Plot of the deviation of $LE$ from $1$ versus
$\ep$ for Fibonacci driving with $J_1 = 1.1, ~J_2 = 1.5$,
and $T = 0.1$. The data is fitted to the form $\alpha 
\ep^2$, where $\alpha$ is found to be $0.30$.}
\label{fig11} \end{figure}

We have seen in Fig.~\ref{fig8} (a) that the $LE$
saturates at some value after a long number of drives
($n \sim 5000$). We will now analyze this in more detail.
Let us denote the saturation value of $LE$ as $LE_{sat}$.
Clearly, $LE_{sat}$ deviates from 1 because $\ep \ne 0$,
since $\ep$ is the parameter which makes the
unitaries $U_1$ and $U_2$ different from each other. As 
$\ep$ is varied, keeping $T$ fixed at a small
value, we find that $LE_{sat}$ also changes.
Denoting $d= 1 - LE_{sat}$, we find that $d$ scales 
with $\ep$ as $\alpha \ep^2$. This is shown in Table 3,
and a numerical fitting
shows that $\alpha = 0.30$ for the system parameters 
$J_1 = 1.1, J_2 = 1.5$, and $T = 0.1$; see 
Fig.~\ref{fig11}. We can understand this scaling as 
follows. As the unitary changes
back and forth between $U_1$ and $U_2$, the value of 
$J_1/J_2$ changes between $\lam_1 = 1.1/1.5$ and $
\lam_2 = 1.1/(1.5+\ep)$. For small $\ep$, 
Eq.~\eqref{over2} implies that the overlap between 
the end mode wave functions
deviates from 1 by an amount which scales as $\ep^2$.
We therefore expect the $LE$ also to eventually settle
down to a value which differs from 1 by a term of
order $\ep^2$.

In contrast, we find that $LE_{sat}$ does not change significantly if we vary $T$, keeping $\ep$ fixed at a 
small value and $\ep T$ remaining much smaller than 1. 
This is because the overlap between the
left-localized end modes of $U_1$ and $U_2$ depends only
on $\ep$ and not on $T$. Varying $T$ therefore
only rescales the time after which the $LE$ saturates,
but the value of $LE_{sat}$ does not change appreciably.

\subsection{Variation of the long-time behavior of 
the $LE$ with $T$}
\label{sec4d}

\begin{figure}
\centering
\subfigure[]{
\includegraphics[scale=0.48]{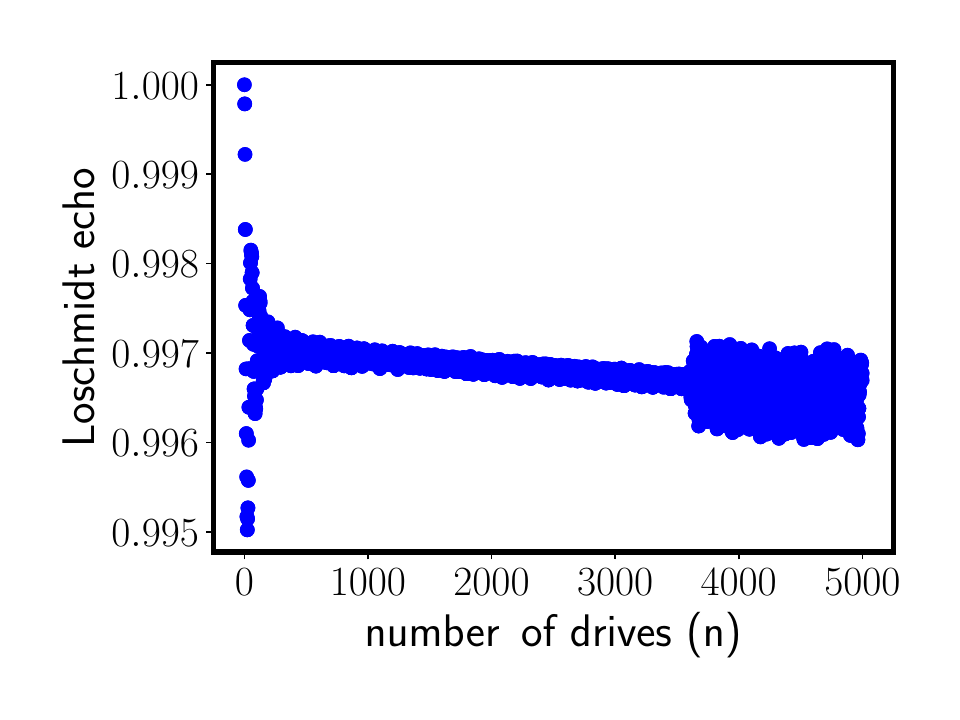}
}
\hfill
\subfigure[]{
\includegraphics[scale=0.48]{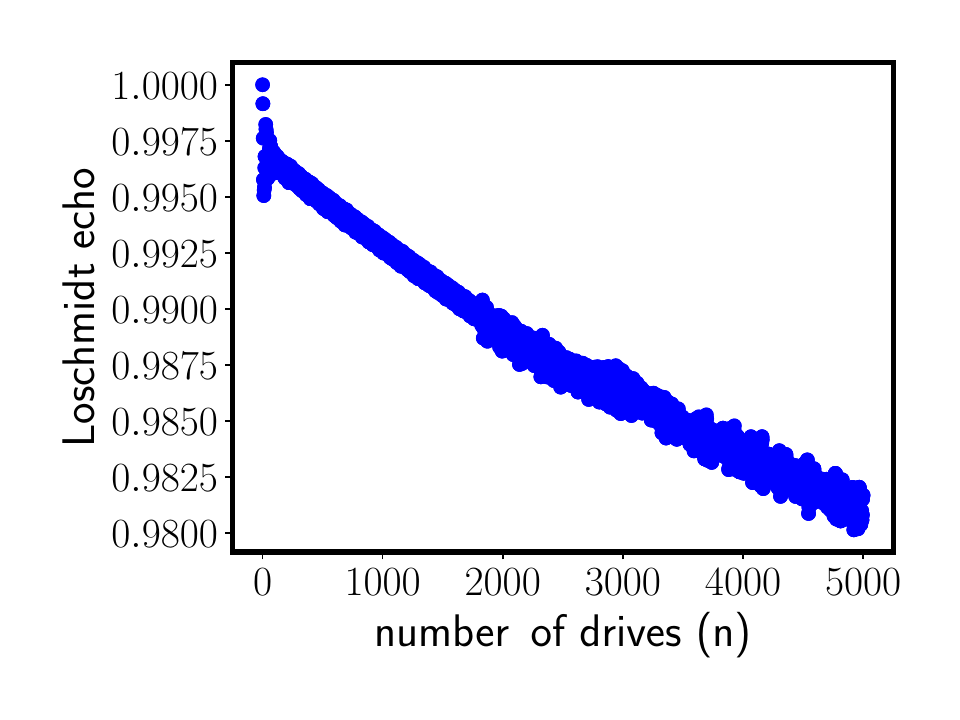}
}
\caption{Plots of the $LE$ versus the number of drives for
Fibonacci driving with (a) $T = 0.2$ and (b) $T = 0.4$, for a system with
$J_1 = 1.1, ~J_2 = 1.5, ~\ep=0.1$, and $L = 800$. The $LE$ 
is seen to decay much faster compared to Fig.~\ref{fig8} (a) 
where $T=0.1$.} \label{fig12} \end{figure}

We have seen in Sec.~\ref{sec4a} that the $LE$ starting from an end mode
remains close to 1 up to a very large number of
drives, up to $n = 10^7$ for the parameters chosen in
Fig.~\ref{fig9}. However, this only holds for small
values of the driving
time period such as $T=0.1$. If $T$ is increased, the $LE$
decays rapidly. This is shown in Fig.~\ref{fig12} for
$T = 0.2$ and $0.4$.

This change of the long-time behavior of the $LE$ as
$T$ is increased can be qualitatively understood as follows. Given
two unitaries $U_1 = e^{-i H_1 T}$ and $U_2= e^{-i H_1 T}$,
we know that their product is given by the 
Baker-Campbell-Hausdorff formula and takes the form
\beq U_1 U_2 ~=~ e^{- i (H_1 + H_2) T ~-~ \frac{1}{2} ~
[H_1, H_2] T^2 ~+~ \cdots}. \label{bch} \eeq
Now, if $H_1$ and $H_2$ differ from each
other by a small amount proportional to $\ep$, the
commutator $[H_1, H_2]$ will be of order $\ep$. In addition,
if $T$ is also small, it is clear the commutator term in
Eq.~\eqref{bch} will be of order $\ep T^2$ which is smaller
than the first term, $(H_1 + H_2 )T$, by a factor of $\ep T$.
Hence, if $\ep T$ is small, we can ignore the commutator
term with respect to the first term. Under this assumption,
the product of a long string of $U_1$'s and $U_2$'s 
will be approximately equal to 
\beq U ~=~ e^{-i (n_1 H_1 + n_2 H_2) T}, \label{bch2} \eeq
where $n_1$ and $n_2$ denote the number of appearances
$U_1$ and $U_2$ respectively. For a very long Fibonacci
sequence of length $n$, it is known that~\cite{maity19}
\bea \frac{n_1}{n} &\to& b ~\simeq~ 0.618, \non \\
\frac{n_2}{n} &\to& 1 - b ~\simeq~ 0.382, \label{gold2}
\eea
where $b$ is given in Eq.~\eqref{gold}. We therefore obtain
\beq U ~=~ e^{-i [b H_1 + (1-b) H_2] n T} \label{bch3} \eeq
when $n$ is large. Defining the effective Hamiltonian $H_{eff}$ through the relation $U = 
e^{-i H_F n T}$, we see that
\beq H_{eff} ~=~ b H_1 ~+~ (1-b) H_2. \label{heff} \eeq
For the system parameters $J_1 = 1.1$, $J_2 = 1.5$ and
$\ep = 0.1$, we see that $H_1$ and $H_2$ with staggered
hopping amplitudes $(1.1,1.5)$ and $(1.1,1.6)$ are both in a
topological phase, and
$H_{eff}$ in Eq.~\eqref{heff} is also in a topological
phase. Hence all three Hamiltonians host zero-energy 
end modes; furthermore, the small value of $\ep$
implies that the overlaps of all the left-localized 
end modes are close to 1. As a result, when the unitary
flips back and forth between $U_1$ and $U_2$, the overlap
with any of the end modes stays close to 1.

The above arguments break down if $\ep T$ 
is not very small or if $n$ is extremely large. Then the commutators appearing in Eq.~\eqref{bch}
become important. For a long Fibonacci sequence of
$U_1$'s and $U_2$'s, it is known that inclusion of all the 
first-order commutators modifies Eq.~\eqref{bch3} to
\beq U ~=~ e^{-i \{b H_1 + (1-b) H_2 - \frac{i}{2}
[H_1, H_2] T \delta (n)\} nT}, \label{bch4} \eeq
where $\delta (n)$ fluctuates rapidly with $n$ but 
always remains within
the range $[-b,1-b] \simeq [-0.618,0.382]$ (see 
Supplemental Material of Ref.~\cite{maity19}). 
The presence of $[H_1, H_2] T$ means that the fluctuating 
term is about $\ep T$ times the first two terms, 
$b H_1 + (1-b) H_2$, and is therefore negligible if 
$\ep T$ is very small. But if $\ep T$ is not so small, 
the fluctuations become significant and lead to a rapid
decrease in the value of the $LE$. This explains the 
plots for the $LE$ versus $n$ shown in Fig.~\ref{fig12}.

In this section, we have only considered the first-order
commutators, namely, $[H_1,H_2]$. As $n$ increases, terms like
$[H_1,[H_1,H_2]]$ and higher-order commutators will become 
increasingly more important. It is possible that
these terms will eventually lead to the
decay of the $LE$ for extremely large values of $n$ as shown
in Fig.~\ref{fig9}.

\subsection{Variation of the time $T_p$ with $T$ and $J_1$}
\label{sec4e}

To quantify the stability of the $LE$, it is useful to
define a time $T_p = nT$ (where $n$ denotes the number of
drives) at which large oscillations begin in the
$LE$. In this section, we will examine how $T_p$ varies 
with the time period $T$ and the hopping $J_1$, keeping
$J_2$ fixed.

\begin{figure}
\centering
\includegraphics[scale=0.48]{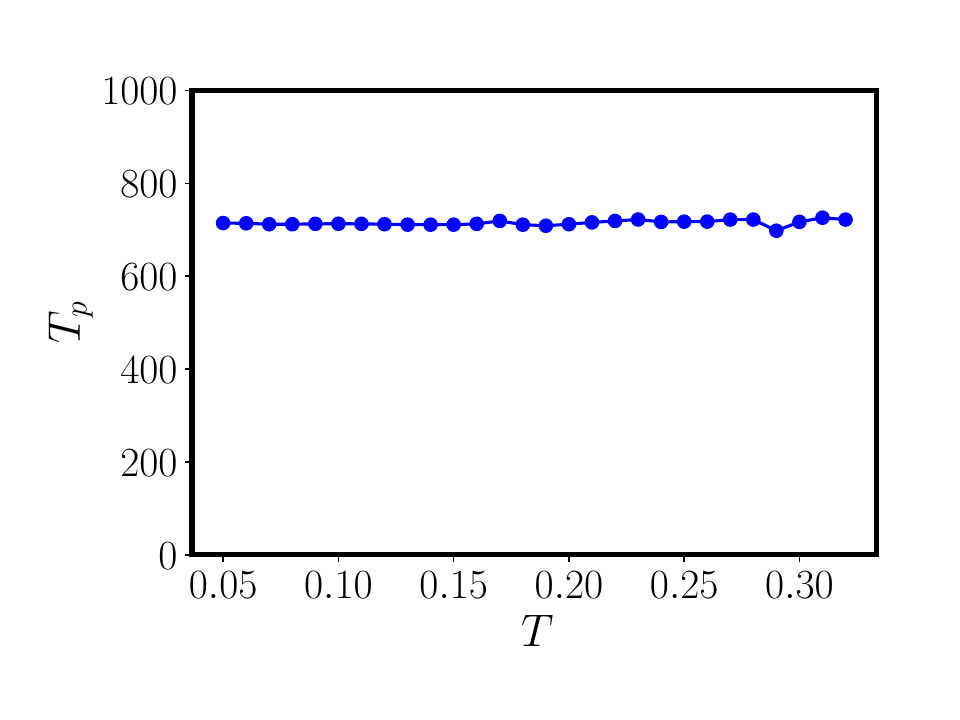}
\caption{Plot of $T_p$ versus $T$ for Fibonacci driving
with fixed $\ep = 0.1$.
We find that $T_p$ remains at almost the same value
of about 700. The system parameters are $J_1 = 1.1, ~J_2 = 1.5$, and $L = 800$.} \label{fig13} \end{figure} 

The variation of $T_p$ with $T$ is shown in Fig.~\ref{fig13}
for $J_1 = 1.1, ~J_2 = 1.5, ~\ep = 0.1$ and $L=800$. We see
that $T_p$ is independent of $T$. This can be understood from 
the discussion in Sec.~\ref{sec4d}. When $\ep T \ll 1$,
the product of a long string of $U_1$'s and $U_2$'s is
approximately given by Eq.~\eqref{bch3} since the commutator
terms can be ignored. This implies that 
$U$ (and hence the $LE$) only depends on the total time 
$t=nT$ and not on $n$ and $T$ separately. Hence the 
stability time $T_p$ is independent of $T$.

\begin{figure}[h]
\centering
\includegraphics[width=0.9\linewidth]{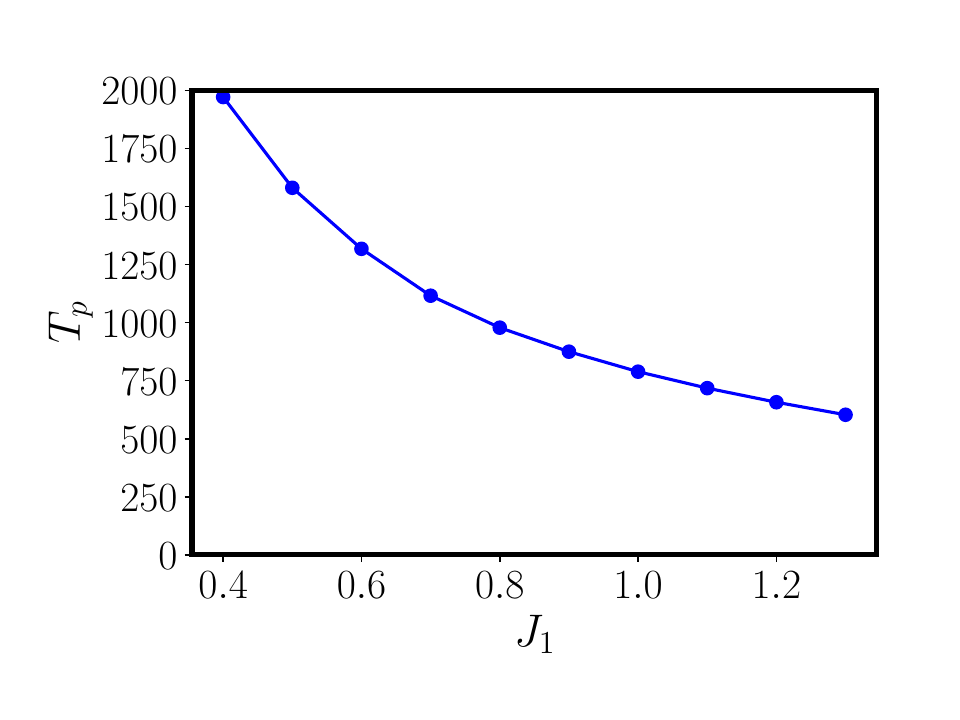}
\caption{Plot of $T_p$ versus $J_1$ for Fibonacci driving. 
The system parameters are $J_2 = 1.5, ~\ep = 0.1$,
$T=0.1$, and $L = 800$.} \label{fig14} \end{figure}

In contrast, we find that $T_p$
varies significantly with the hopping $J_1$, as
shown in Fig.~\ref{fig14}. We find that $T_p$ increases as
$J_1$ decreases and vice versa. This can be understood as
follows. Equation~\eqref{endmode} implies that the decay length 
of the end modes, given by $\xi$ where $e^{-1/\xi} = 
J_1/J_2$, decreases as $J_1 / J_2 \to 0$. Hence the 
left-localized end mode 
becomes more localized at the leftmost
site of the system as $J_1$ decreases; as a result it
becomes increasingly immune to small changes in $J_2$
which occurs due to the unitary changing back and forth
between $U_1$ and $U_2$. The $LE$ therefore remains close
to 1 for a long time, leading to larger values of $T_p$.
In contrast, as $J_1$ approaches $J_2$, the decay length
of the end mode becomes large and also varies more as 
$J_2$ changes by small amounts. As a result, the end mode 
becomes more unstable which leads to smaller values of $T_p$.

\section{Thue-Morse driving with two unitaries}
\label{sec5}

Given two unitaries $U_1$ and $U_2$, we can
generate an aperiodic Thue-Morse sequence as follows~\cite{nandy17}.
We define $A_1 = U_1$ and $B_1 = U_2$, and then
recursively define $A_{n+1} = B_n A_n$ and $B_{n+1}
= A_n B_n$ for $n \ge 2$. These sequences have a 
length with grows exponentially as $2^n$. The first 
few sequences $A_n$ are given by 
\begin{eqnarray}
A_1 &=& U_1, \quad A_2= U_2 U_1, \quad A_3= U_1 U_2 U_2 U_1, \non \\
A_4 &=& U_2 U_1 U_1 U_2 U_1 U_2 U_2 U_1, \label{tms}
\end{eqnarray}
and so on. We then define the 
Thue-Morse driving as the one obtained by acting with
$A_1, ~A_2, ~\cdots$ on an initial state $| \psi_1 \ra$.

\begin{figure}
\centering
\includegraphics[scale=0.48]{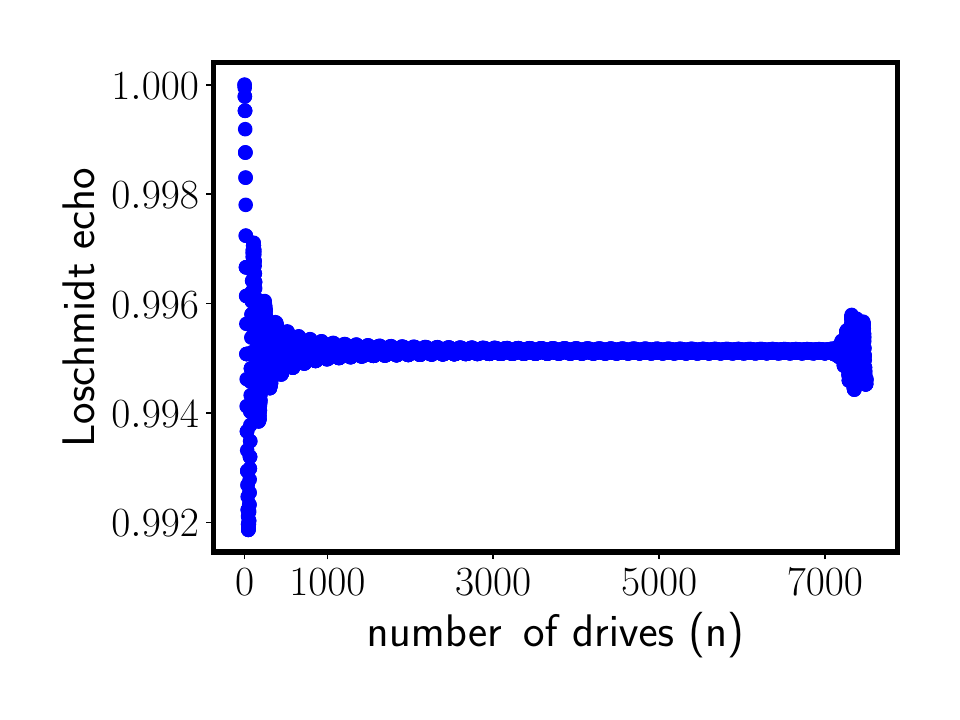}
\caption{Plot of the $LE$ versus the number of
drives for Thue-Morse driving of a system with
$J_1 = 1.1, ~J_2 = 1.5, ~\ep=0.1, ~T=0.1$, and $L = 800$. 
The initial state is taken to the left-localized end mode
of $U_1$.} \label{fig15} \end{figure}

\begin{figure}
\centering
\subfigure[]{
\includegraphics[scale=0.48]{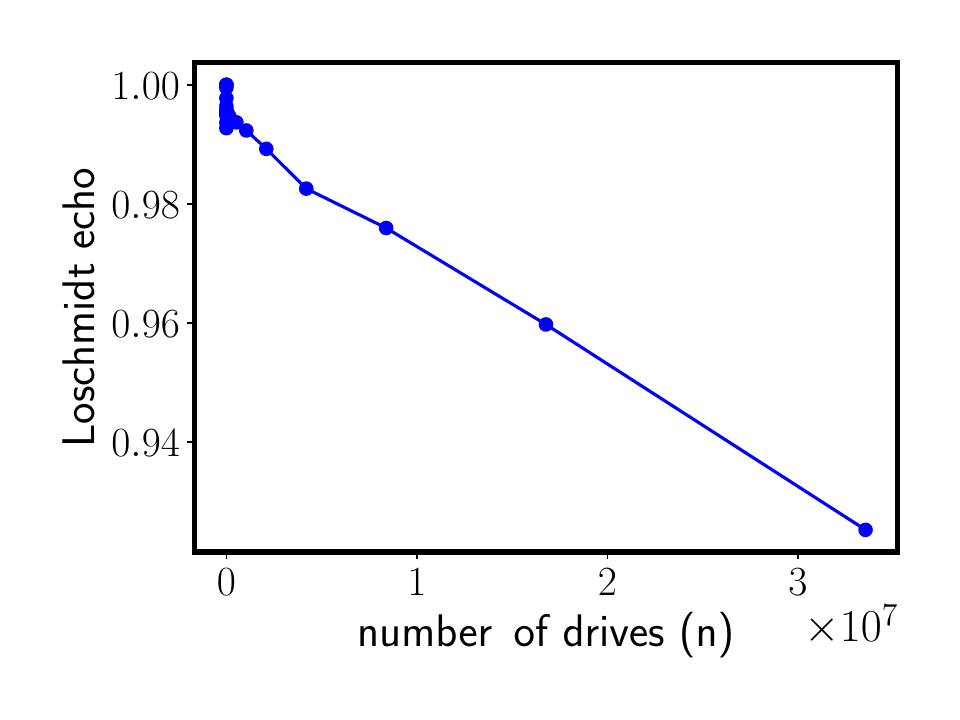}
}
\hfill
\subfigure[]{
\includegraphics[scale=0.48]{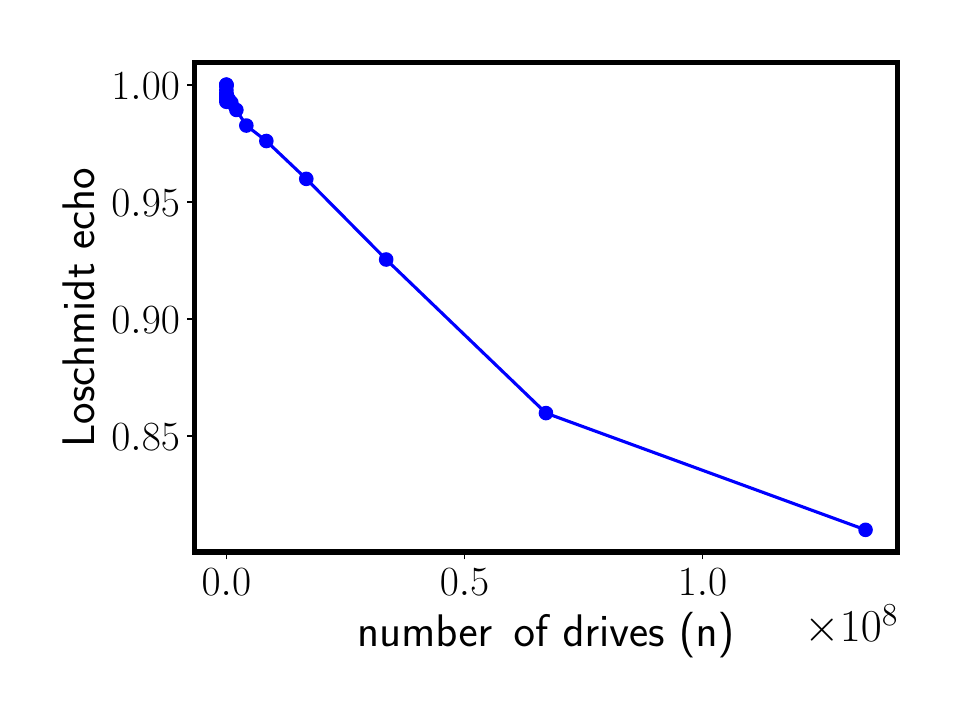}
}
\caption{Plots of the $LE$ versus the number of drives for 
Thue-Morse driving of a system with
$J_1 = 1.1, ~J_2 = 1.5, ~\ep=0.1, ~T=0.1$, and $L = 800$. 
The number of drives is exponentially large, going up to
about $3.2 \times 10^7$ in (a) and $1.2 \times 10^8$ in (b).
The decay rate is comparable to what we see for Fibonacci
driving in Fig.~\ref{fig9}.} \label{fig16} \end{figure}

Alternatively, we can generate Thue-Morse sequences 
whose lengths grow linearly as follows. Given an 
integer $n \ge 1$, we first write $n-1$ as a binary 
number $b_n$. Thus, the numbers $n~=~1,~2,~3,~4,~5,~6,~7,~8,~
\cdots$ lead to the binary numbers $b_1=0, ~b_2=1, ~b_3=10, ~b_4 = 11, ~b_5=100, ~b_6=101, ~b_7=110, ~b_8=111,~\cdots$. 
Next, for a number $b_n$, we add up all the digits;
if the sum is equal to 0 mod 2 we map $n$ to $U_1$, 
whereas if
the sum is equal to 1 mod 2 we map $n$ to $U_2$.
We see that the numbers $1,~2,~\cdots,~8$ map to
$U_1, ~U_2, ~U_2, ~U_1, ~U_2, ~U_1, ~U_1, ~U_2$;
putting these together from right to left generates
the sequence $A_4$ in Eq.~\eqref{tms}.

For Thue-Morse driving, we can understand the stability of 
the $LE$ up to quite
large values of $n$ (see Fig.~\ref{fig15}) using ideas 
similar to those presented in Sec.~\ref{sec4d} for Fibonacci 
driving. For $\ep T \ll 1$, we can ignore the contribution 
of the commutator $[H_1,H_2]$ and obtain an
expression for the operator $U$ after a large number of
drives similar to Eq.~\eqref{bch3}. For Thue-Morse driving, 
we find that
\beq U ~=~ e^{- (i/2) [H_1 + H_2] n T} \label{bch5} \eeq
when $n$ is large, since $U_1$ and $U_2$ appear
an equal number of times on the average. Since the overlaps between the
end modes of $H_1$, $H_2$ and $(H_1+H_2)/2$ are all 
close to 1 if $\ep$ is small, the $LE$ will also stay 
close to 1 for quite large values of $n$. However, if
$n$ is extremely large,
commutators of different orders will begin to
contribute significantly even if $\ep T \ll 1$,
and the $LE$ will deviate from 1 as we see in 
Fig.~\ref{fig16}. This can again be understood by 
arguments similar to those presented around Eq.~\eqref{bch4}. 
Namely, if $n$ is extremely large or if $\ep T$ is not
very small, the commutators appearing in Eq.~\eqref{bch}
become important. For a long Thue-Morse sequence of
$U_1$'s and $U_2$'s, it can be shown that including all the first-order commutators modifies Eq.~\eqref{bch5} to
\beq U ~=~ e^{-i \{ (H_1 + H_2)/2 - \frac{i}{2}
[H_1, H_2] T \delta (n)\} nT}, \label{bch6} \eeq
where $\delta (n)$ fluctuates rapidly with $n$ but 
always remains within the range $[-0.5,0.5]$. 
The presence of $[H_1, H_2] T$ means that the fluctuating 
term is about $\ep T$ times the first two terms, 
$(H_1 + H_2)/2$, and is therefore negligible if 
$\ep T$ is very small. But if $\ep T$ is not so small, 
the fluctuations become significant and lead to a rapid
decrease in the value of the $LE$.

\section{Random driving with two unitaries}
\label{sec6}

\begin{figure}[h]
\centering
\subfigure[]{
\includegraphics[scale=0.48]{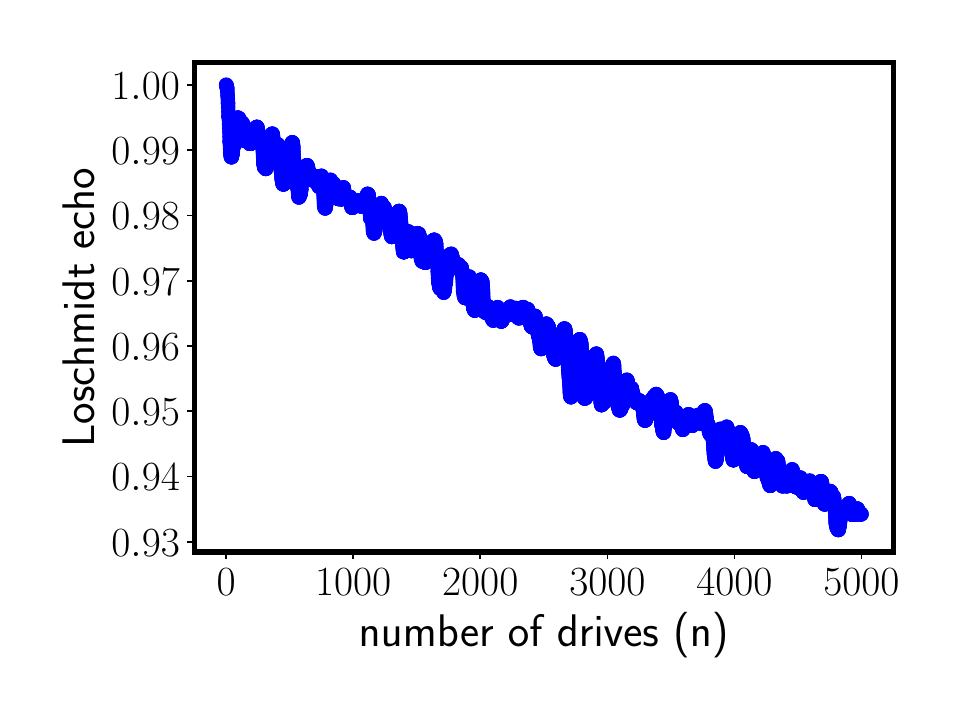}
}
\hfill
\subfigure[]{
\includegraphics[scale=0.48]{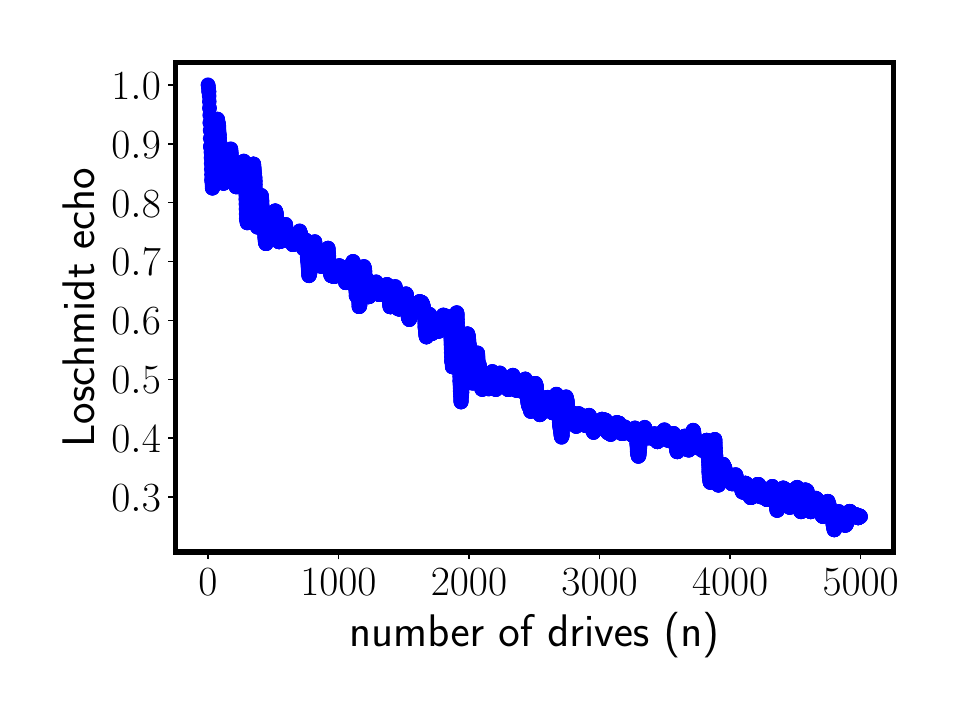}
}
\caption{Plots of the $LE$ versus the number of drives $n$
for a random protocol, for (a) $\ep = 0.1$ and (b) $\ep = 
0.5$, for a system with $J_1 = 1.1,~ J_2 = 1.5$, $T = 0.1$, 
and $L=800$. We see that the $LE$ decays very rapidly 
compared to what is seen for a Fibonacci drive in 
Fig.~\ref{fig9}.} \label{fig17} \end{figure}

To contrast with the special features of the
Fibonacci drive, we now consider driving the same 
left-localized end mode of $H_1$ with a 
random drive for parameters $J_1 = 1.1, ~J_2 = 1.5$, $T = 0.1$, $L=800$, and $\ep = 0.1$ and $0.5$. Figure~\ref{fig17}
shows plots of the $LE$ versus the number of drives $n$
for the two values of $\ep$. For both values of $\ep$, we find
that the $LE$ starts decaying quite early (about $n \sim 20$),
and it decays much faster than for the
Fibonacci drive (Fig.~\ref{fig9}). Thus
the end modes remain stable up to a much larger number of
drives ($n \sim 10^7$) for the Fibonacci and Thue-Morse 
drives compared to a random drive where the end modes become
unstable quite rapidly. This can be understood by looking 
at the effects of first-order commutators following arguments similar
to those given above for the Fibonacci
and Thue-Morse drives. For a random drive we find that 
\beq U ~=~ e^{-i \{ (H_1 + H_2)/2 - \frac{i}{2}
[H_1, H_2] T \delta (n)\} nT}, \label{bch7} \eeq
where the coefficient of $H_1 + H_2$ 
is $1/2$ because $U_1$ and $U_2$ appear with probability
$1/2$ each. However, unlike the expressions given in
Eq.~\eqref{bch4} and \eqref{bch6} for the Fibonacci and
Thue-Morse drives respectively, we find that for a 
random drive, the
rapidly fluctuating number $\delta (n)$ in Eq.~\eqref{bch7}
does {\it not} lie in a bounded interval but rather grows
as $\sqrt{n}$ along with a fluctuating sign. (This is essentially because a random walk consisting
of two possible steps, $+1$ and $-1$, will show fluctuations
of order $\sqrt{n}$ after $n$ steps). As a result, the
effective Hamiltonian $H_{eff}$ obtained from 
Eq.~\eqref{bch7} by $U = e^{-i H_{eff} nT}$
fluctuates in an unbounded manner when $n$ becomes
very large. This leads to a rapid decay in the $LE$.

\begin{figure}[h]
\centering
\includegraphics[width=0.95\linewidth]{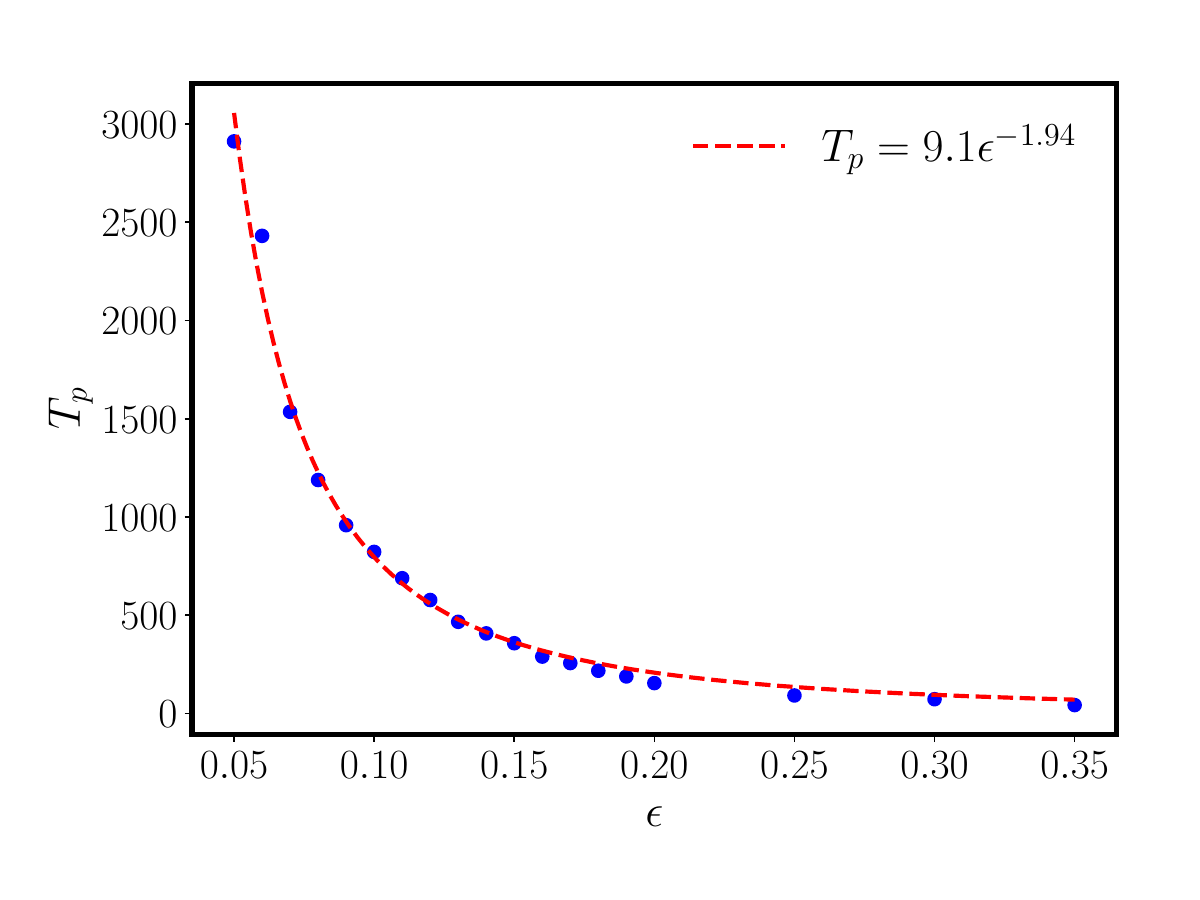}
\caption{Plot of the time $T_p$ at which the $LE$ for a
random protocol decreases to the value of $0.9$ versus $\ep$,
for a system with $J_1 = 1.1, ~J_2 = 1.5$, $T = 0.1$, 
and $L=800$.} \label{fig18} \end{figure}

To quantify how fast the $LE$ decays for a random protocol,
we define a time $T_p = n T$ where the $LE$ decreases 
from 1 to a value of $0.9$. Figure~\ref{fig18} shows $T_p$
versus $\ep$ for the same system parameters as in 
Fig.~\ref{fig17}. A power-law fitting shows that $T_p$ scales 
as $1/\ep^{1.94}$ which is fairly close to $1/\ep^2$. We can
understand this as follows. As the unitary randomly changes
back and forth between $U_1$ and $U_2$, the value of 
$J_1/J_2$ changes between $\lam_1 = 1.1/1.5$ and $
\lam_2 = 1.1/(1.5+\ep)$.
We saw earlier that, for small $\ep$,
the overlap between the end mode wave functions
deviates from 1 by an amount which scales as $\ep^2$.
We therefore expect that each time the unitary randomly
changes between
$U_1$ and $U_2$, the $LE$ should decrease by a factor which
deviates from 1 by a term of order $\ep^2$; let us write this
factor as $e^{-\ga \ep^2}$, where $\ga$ is a number of order 1. 
For a random sequence of $U_1$'s and $U_2$'s with a large 
length $n$, it is known that the number
of times the unitary operator changes between $U_1$ and 
$U_2$ is, on the average, given by $n/2$. (This is simply because after each unitary $U_i$, where $i$ can be 1 or 2, 
the probability that the next unitary will be the opposite 
of $U_i$ is $1/2$). We therefore expect that
after $n$ drives, the $LE$ will decrease to a value of about
$e^{- (n \ga/2) \ep^2}$. This will be equal to $0.9$ when
\beq n ~=~ - ~\frac{2 \ln (0.9)}{\ga \ep^2}. \eeq
Hence $T_p = n T$ should scale as $1/\ep^2$.

\vspace*{-.4cm}
\section{Discussion}
\label{sec7}

We now summarize our results. We have 
considered the SSH model in which there are staggered 
hopping amplitudes $J_1$ and $J_2$. For an open chain with
the leftmost and rightmost bonds have a hopping amplitude
$J_1$ which is smaller than $J_2$, it is well known,
as mentioned in Sec.~\ref{sec2}, that 
there is one zero-energy topological mode 
localized at each end of the system. 

In Sec.~\ref{sec3}, we study
what happens if the hopping $J_2$ of the SSH
model is varied periodically
in time. For simplicity, we take the periodic 
variation to have the form of a square pulse, so that
$J_2$ alternates between two values. We denote the
corresponding Floquet evolution operators as $U_1$ and 
$U_2$. The combination of the two is given by $U= U_2 U_1$. Depending on the system parameters, we find 
that driving by the operator $U$ can generate multiple 
modes at each end of the system. These modes may
be topological; in that case the number of modes
at each end agrees with the winding number 
which is a topological invariant, and the Floquet
eigenvalues of these modes are exactly equal to 
$+1$ or $-1$ for a sufficiently large system
size. But if the end modes are non-topological,
their number does not agree with the winding number, and
their Floquet eigenvalues are not equal to $\pm 1$.
We see examples of both topological and non-topological end
modes depending on the driving parameters. Next, we study 
what happens if an end mode of $H_1$ is acted upon repeatedly by $U$.
We find that the Loschmidt amplitude $LA$, defined as the  overlap between
the initial state and the state obtained after $n$ drives,
shows pronounced oscillations. The Fourier transform
of the $LA$ has a sharp peak at a frequency $\Omega$
which is equal to the quasienergy of the end-localized
mode of $U$ with which the initial state has the largest
overlap.

We then study in Sec.~\ref{sec4} what happens if an end mode 
is acted upon by a Fibonacci sequence of two 
unitaries $U_1$ and $U_2$ which are close to each other. Namely,
the values of $J_2$ for $U_1$ and $U_2$ differ from each
other by a small amount called $\ep$, and both $U_1$ and
$U_2$ are taken to act for a time $T$ which is also small.
Depending on the quantity of interest, we have taken the 
length of the driving
sequence to grow either linearly or exponentially
as the Fibonacci numbers. When both $\ep T \ll 1$, 
we find that the $LE$ (defined as the modulus squared of the $LA$) oscillates about a mean value which is 
close to 1 for a very large number of drives $n$, which is of the
order of $10^7$ for our choice of parameters. However, if
$n$ is very large, of the order of $10^8$ or more, the $LE$
starts deviating substantially from 1. In contrast, if
$\ep T$ is not much smaller than 1, the $LE$ starts deviating
from 1 quite rapidly. We provide an understanding of this
difference between the behaviors of the $LE$ for $\ep T \ll 1$
and $\ep T \sim 1$ based on the Baker-Campbell-Hausdorff
formula for the product of a large of unitaries. We also
study how the behavior of the $LE$ for large $n$ 
depends on the parameter $J_1$ (we always assume that
$J_1 < J_2$ so that end modes exist). We find that the $LE$
stays close to 1 for $J_1 \ll J_2$ but deviates quickly from 1
as $J_1$ approaches $J_2$. This difference can be understood
based on the fact that the end modes are highly localized and mix
very little with the bulk modes $J_1 \ll J_2$, whereas the end
modes have a large decay length and quickly mix with the bulk
modes when $J_1$ approaches $J_2$.
In Sec.~\ref{sec5} we have examined what
happens when an aperiodic sequence, called the 
Thue-Morse sequence, of $U_1$ and $U_2$'s act on an end mode.
We find that the effects of Fibonacci and Thue-Morse sequences
of drives are quite similar~\cite{ghosh25}. This stems
from the recursive nature of these sequences which leads to 
significant cancellations in the first-order commutators 
which appear in the Baker-Campbell-Hausdorff formula.

In Sec.~\ref{sec6}, we have studied what happens 
when a random sequence
of $U_1$ and $U_2$'s act on an end mode. In this case, we
find that the $LE$ decays from 1 quite quickly, even when
$\ep$ and $T$ are small. For very small values of $\ep$, 
we find that the decay time $T_p$ of the $LE$ scales 
approximately as $1/\ep^2$. This can be understood as arising
from the fact that the overlap of the end modes of $U_1$ and $U_2$ differs from 1 by an amount which scales as $\ep^2$.
A random sequence of $U_1$ and $U_2$'s therefore degrades the
state by an amount which differs from 1 by a term of order 
$\ep^2$.

We end by pointing out some directions for future 
studies. For quasiperiodic and aperiodic driving 
(Fibonacci and Thue-Morse respectively), we saw 
that when $\ep$ and $T$
are both small, the $LE$ is stable up to quite large 
values of the drive number $n$, of the order of $5000$,
while for extremely large values of $n$, of the order
of $10^8$, the $LE$ starts deviating appreciably from 1.
We have mentioned in Sec.~\ref{sec4d} that this may be 
due to higher commutators, but this needs to be
understood in detail. Next, we note that with periodic
boundary conditions, the system decouples into a product
of two-level systems labeled by a momentum $k$. For a
two-level system, both Fibonacci and Thue-Morse driving 
are known to lead
to some conserved quantities~\cite{nandy17,nandy18}. It may
be interesting to study the consequences of these conservation
laws for the long-time dynamics. Finally, it may be
instructive to investigate how the phase diagram of the SSH model
evolves with time as a function of the different parameters 
for quasiperiodic driving. In particular, one can study if
the phase diagram, found via the boundary autocorrelation 
function, shows a self-similar structure as is known
to occur for Fibonacci driving of the transverse field 
Ising model~\cite{schmid25}.

Finally, we note that SSH systems with Fibonacci 
quasiperiodic sequences of on-site potentials or 
nearest-neighbor hoppings have been studied
earlier from different points of view. For 
instance, see Ref.~\cite{ouyang24} for a study of 
higher-order topology in such systems, and Refs.~\cite{mukherjee1,mukherjee2} for experimental
realizations in photonic lattices. It would be interesting to study the dynamics when such systems are
subjected to the various driving protocols discussed here.

\vspace*{0.5cm}
\centerline{\bf Acknowledgment}
\vspace{0.4cm}

D.S. thanks SERB, India for funding through Project No. JBR/2020/000043.

\end{document}